\documentclass[reprint,aps,prl,balancelastpage,a4paper,floatfix,superscriptaddress]{revtex4-2}
\usepackage[margin=2cm]{geometry}
\usepackage[T1]{fontenc}
\usepackage{graphicx}
\usepackage{float}
\usepackage{amsmath,amssymb,bm}
\usepackage{xcolor}
\usepackage{booktabs}
\usepackage{enumitem}
\usepackage{dcolumn}
\usepackage{hyperref}
\usepackage{hycolor}
\hypersetup{colorlinks=true,citecolor=[rgb]{0,0.15,0.60}, 
urlcolor=[rgb]{0,0.15,0.60}, linkcolor=[rgb]{0,0.15,0.60}, final = true}
\usepackage[rightcaption]{sidecap}
\usepackage{physics}
\usepackage{mathtools}
\usepackage{relsize}
\usepackage{times}
\usepackage{siunitx}
\setcitestyle{round,super}

\def\JPE#1{JPE: \color{red} {\tt #1}}
\def\JPE#1{\kern 30ptJPE:\ \color{red}{ #1}\color{black}}
\def\T{{\rm T}}
\def\be{\begin{equation*}}
\def\ee{\end{equation*}}
\def\ba{\begin{eqnarray}}
\def\ea{\end{eqnarray}}
\def\eref#1{(\ref{#1})}
\def\eref#1{Eq.(\ref{#1})}
\def\fref#1{Fig.~\ref{#1}}

\def\CC{{\mathcal{C}}}
\def\DD{{\mathcal{D}}}
\def\MM{{\mathcal{M}}}

\def\PP{{\mathcal{P}}}
\def\QQ{{\mathcal{Q}}}
\def\EE{{\mathcal{E}}}

\def\ZZ{{\mathcal{Z}}}
\def\Hb{{\mathbf{H}}}
\def\rb{{\mathbf{r}}}
\let\epsilon=\varepsilon
\def\l{\left}
\def\r{\right}
\def\DG{\upDelta G}
\def\DS{\upDelta S}

\def\Dr{\Delta r}
\def\Keff{K_\mathrm{eff}}

\def\As{{\mathsmaller{\mathrm{A}}}}
\def\Bs{{\mathsmaller{\mathrm{B}}}}
\def\Ss{{\mathsmaller{\mathrm{S}}}}
\def\Cs{{\mathsmaller{\mathrm{C}}}}
\def\NCs{{\mathsmaller{\mathrm{NC}}}}
\def\Ss{{\mathsmaller{\mathrm{S}}}}
\def\Ls{{\mathsmaller{\mathrm{L}}}}
\def\Lams{{\mathsmaller{\mathrm{\Lambdaup}}}}

\def\DGC{\DG_\Cs}
\def\DGNC{\DG_\NCs}
\def\DGL{\DG_\Ls}
\def\DGS{\DG_\Ss}
\def\DSC{\upDelta S_{\mathrm{conf}}}
\def\DSO{\upDelta S_0}
\def\bDG{\upDelta\bar{G}}
\def\KL{K_\Lams}
\def\Ks{K_\mathrm{s}}
\def\Km{K_\mathrm{m}}
\def\Kw{K_\mathrm{w}}
\def\epss{\epsilon_\mathrm{s}}
\def\epsw{\epsilon_\mathrm{w}}

\def\NA{{n_\As}}
\def\NB{{n_\Bs}}
\def\NS{{n_\Ss}}

\def\thetL{\theta_\Ls}
\def\thetS{\theta_\Ss}
\def\thetC{\theta_\Cs}
\def\thetNC{\theta_\NCs}
\def\half{{\textstyle\frac{1}{2}}}
\def\sb#1{\textbf{\textsf{#1}}}
\def\nsb#1{\noindent\textbf{\textsf{#1.~}}}

\usepackage{amstext}
\usepackage[normalem]{ulem}

\DeclareMathSymbol{:}{\mathord}{operators}{"3A}

  \definecolor{or}{RGB}{234,142,53}
  \definecolor{gr}{RGB}{150,150,150}
  \definecolor{bl}{RGB}{54,152,187}
  \newcommand{\ie}{\textit{i.e.}}
  \newcommand{\eg}{\textit{e.g.}}

\usepackage{mathtools}
\usepackage{libertine}
\usepackage[slantedGreek,libertine]{newtxmath}

\newcommand{\edf}{\EE}
\newcommand{\ech}{\CC}

\newcommand{\DDG}{\upDelta\upDelta G}
\newcommand{\DDGLS}{\DDG_{\Ls\Ss}}
\newcommand{\dlp}{\upDelta \theta_0}

\newcommand{\dlig}{\upDelta \theta_{\Ls\Ss}}

\newcommand{\LW}{\mathbf{w}}
\newcommand{\LS}{\mathbf{s}}
\newcommand{\LLW}{\mathbf{w}_c}
\newcommand{\LLS}{\mathbf{s}_c}

\definecolor{YKB}{rgb}{0.00,0.18,0.65}
\definecolor{YInMn}{rgb}{0.19,0.42,0.75}

\begin{document}

\title{\sb{\Large General theory of specific binding: \\ insights 
from a genetic-mechano-chemical protein model}}

  \author{John M. McBride}
    \email{jmmcbride@protonmail.com}
    \affiliation{Center for Soft and Living Matter, Institute for Basic Science, Ulsan 44919, South Korea}
  \author{Jean-Pierre Eckmann} 
    \email{Jean-Pierre.Eckmann@unige.ch}
    \affiliation{D\'{e}partement de Physique Th\'{e}orique and Section de Math\'{e}matiques, University of Geneva, Geneva, Switzerland}
  \author{Tsvi Tlusty}
    \email{tsvitlusty@gmail.com}
    \affiliation{Center for Soft and Living Matter, Institute for Basic Science, Ulsan 44919, South Korea}
    \affiliation{Departments of Physics and Chemistry, Ulsan National Institute of Science and Technology, Ulsan 44919, South Korea}

\begin{abstract}
  Proteins need to selectively interact with specific targets among a multitude of similar molecules in the cell. But despite a firm physical understanding of binding interactions, we lack a general theory of how proteins evolve high specificity. Here, we present such a model that combines chemistry, mechanics and genetics, and explains how their interplay governs the evolution of specific protein-ligand interactions. The model shows that there are many routes to achieving molecular discrimination -- by varying degrees of flexibility and shape/chemistry complementarity -- but the key ingredient is precision. Harder discrimination tasks require more collective and precise coaction of structure, forces and movements. Proteins can achieve this through correlated mutations extending far from a binding site, which fine-tune the localized interaction with the ligand. Thus, the solution of more complicated tasks is enabled by increasing the protein size, and proteins become more evolvable and robust when they are larger than the bare minimum required for discrimination. The model makes testable, specific predictions about the role of flexibility and shape mismatch in discrimination, and how evolution can independently tune affinity and specificity.
  Thus, the proposed theory of specific binding addresses the natural question of ``why are proteins so big?''. A possible answer is that molecular discrimination is often a hard task best performed by adding more layers to the protein.
\end{abstract}

\maketitle

\section*{\sb{Introduction}}

  Proteins are the main molecular workforce inside cells,
  and the tasks they perform invariably rely on specific, short-range interactions. The cell is filled with thousands of molecular species, some differing by only a single atom. Yet somehow, most proteins can specifically bind to only a few select species.~\cite{namsc12,piace18,coppb20} 
  Despite knowing much about pairwise binding,~\cite{milps97,mccco98,mape99,gohac02,mobst09,wanpn13,kasjr13,weips14,staac16,planc17,wanif20} we understand much less the many-body problem of how proteins \textit{selectively} bind to targets, while avoiding interactions with similar, but non-cognate, molecules.~\cite{kasjr13}
  Such undesirable interactions can, at best, lead to inefficiencies through inhibition,~\cite{goljg44}
  and at worst, result in aggregation~\cite{mahjp09}, inaccurate translation, or cross-talk between signals.~\cite{olssi00} 
  
  Unwanted interactions can be minimized by designing mismatch between ligands and binding pockets,~\cite{rohar10,sernc20} such that the energetic cost of deformation allows the protein to sift the target from similar, non-cognate ligands -- a form of `conformational proofreading'.~\cite{savpo07,savce13}
  Recent work has suggested that residues not directly involved in binding may play a role in discrimination via allostery.~\cite{olipr19,eckrmp19,wanel22,dutpn18} Here, taking inspiration from the many experimental examinations of molecular discrimination by proteins, such as enzymes, tRNA synthetases, transcription factors, and antibodies ,~\cite{sunap02,elina12,namsc12,jannr15,huapn15,piace18,marac18,litel20,bisbs20,tawfj20} we propose a simple yet general theory of specific binding.

  We study the evolution of discrimination by proteins using a genetic-mechano-chemical model of binding.
  We find that, although the discrimination problem is in general difficult, it has many possible solutions: shape mismatch, chemical complementarity, and flexibility can all be manipulated in various ways to tune interaction specificity. The important common thread is that it requires precision and coordination -- for example, just the right amount of shape mismatch to fit a given flexibility. We show that residues distant from the binding site enable this fine-tuning of mechanical deformation upon binding -- demonstrating that discrimination is the outcome of concerted, collective interactions throughout the protein. Thus, larger proteins benefit from having more degrees of freedom, allowing them to solve harder discrimination challenges. Furthermore, larger proteins are more evolvable and robust since their set of possible functional sequences is larger and more connected.
  We further explain the mechanisms through which affinity and specificity can be tuned independently, and discuss the role of flexibility and entropy. Altogether, this simple model shines light on the difficult problem of how proteins achieve such superlative molecular discrimination. At the same time, by linking protein size to a property as ubiquitous as binding, we offer a possible answer to the question, ``why are proteins so big?''.~\cite{paytb83,sretb84}

\section*{\sb{Results}}

  \begin{figure*}[t!]  \centering
  \includegraphics[width=0.99\linewidth]{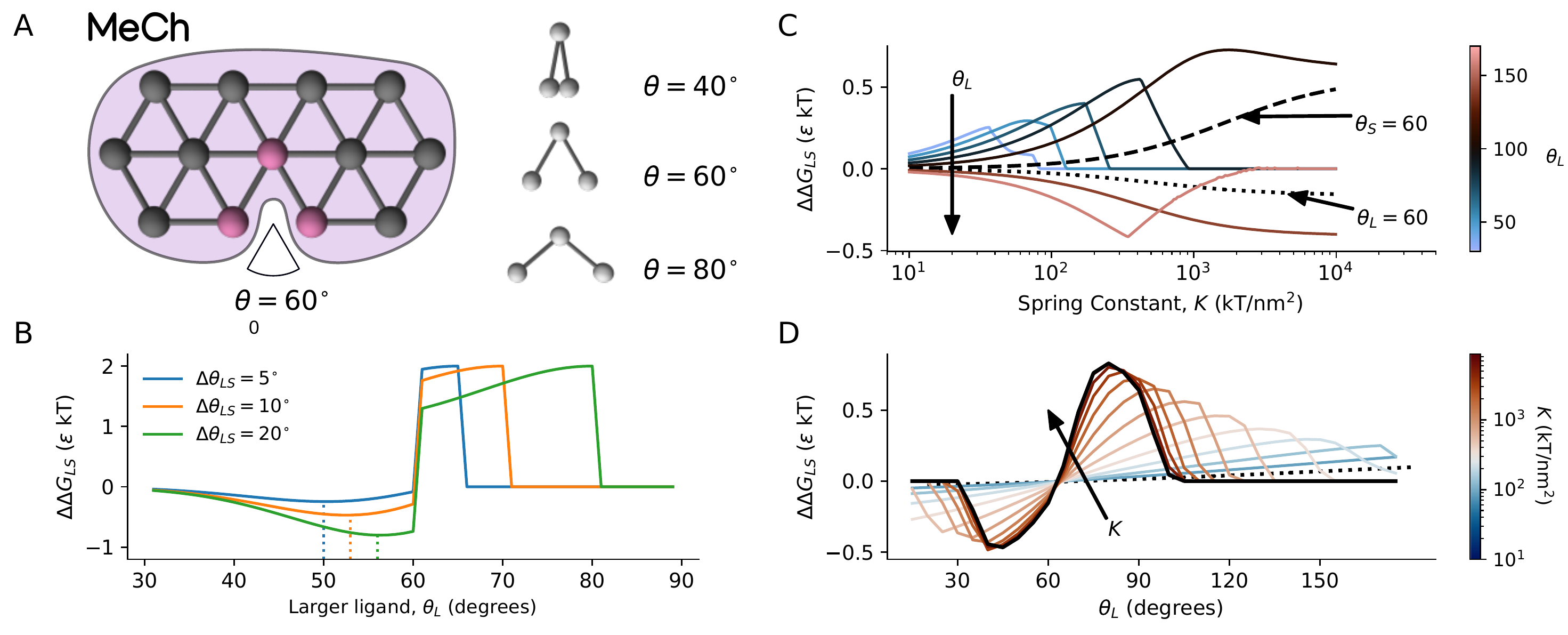}
  \caption{\label{fig:fig1}
  \nsb{Mechano-chemical (\textbf{MeCh}) model of molecular recognition} 
  \sb{A}: A protein is modelled as a 2-d spring network with a chemical binding site (three interaction sites, shown in pink); the $\Lambdaup$-shaped ligands have three interaction sites, and are uniquely defined by an angle $\theta$ (three examples are shown). 
  \sb{B}: Binding free energy gap in a rigid protein $\DDGLS = \DGL - \DGS$ as a function of the size of the larger ligand $\thetL$, for various ligand size differences $\dlig$;
  minima are indicated with dotted lines.
  \sb{C}: $\DDGLS$ as a function of the spring constant $K$ for $\dlig=\ang{5}$, for $\ang{10} \leq \thetL \leq \ang{175}$. 
  \sb{D}: $\DDGLS$ as a function of $\thetL$ for $\dlig=\ang{5}$, for $10\leq K \leq \SI{e4}{kT/\nm^2}$.
  }
  \end{figure*}

\nsb{The models} 
  We introduce three protein models, starting with a bare-bones model and then adding archetypal protein features, so that we achieve a graduated understanding of how proteins   evolve molecular discrimination. We first introduce the basic mechano-chemical protein model (\textbf{MeCh}), a homogeneous spring network with chemical binding sites that bind to a ligand (\fref{fig:fig1}A). This basic model shows how specific binding depends on an interplay of: shape (mis)match between ligands and the protein binding pocket, protein flexibility, chemical binding strength, and entropy. 
  In the second, genetic model (\textbf{G-MeCh}) we introduce a protein sequence which determines the stiffness of individual springs, and the chemical binding strength of binding sites. This allows us to study how specific binding is achieved in a model where discrete changes occur via sequence mutations.
  In the final variant of the model (\textbf{G-MeCh-S}), we allow the equilibrium structure of the protein to change as a function of sequence, and we examine proteins of different sizes. This model facilitates the study of protein evolution, and the effect of protein size, via quantification of evolvability and mutational robustness.\\

\nsb{Mechano-chemical (MeCh) model of molecular recognition} 
  The basic mechano-chemical model is a 2-dimensional elastic network,~\cite{tirpr96,chepb05,lopco16} with $\NA$ amino acids, $a_i$, arranged on a hexagonal lattice, and $\NB=3$ chemical binding sites, each with binding energy constant $\epsilon$ (\fref{fig:fig1}A). 
  In this coarse-grained representation, one may envision the nodes as groups of tightly-connected amino acids that have highly-correlated motion.~\cite{halce09} Likewise, the binding sites can be thought of as a subdivision of a binding site into the three most salient units of amino acids and functional groups.~\cite{ricja19}
  
  The protein has a $\Lambdaup$-shaped binding pocket, described by an angle of $\theta_0 = \ang{60}$; opposite the protein is a set of $\Lambdaup$-shaped ligands with three binding sites, which are uniquely described by an angle in the range $\ang{5} \leq \theta \leq \ang{175}$. Shape mismatch is thus a function of only the ligand shape, and varies along a single dimension, $\dlp = \theta - \theta_0$. 
  Regarding ligands, we use the following terminology: Ligands that do not fit in the binding pocket ($\theta>\ang{60}$) are called `fat', otherwise `thin'. When comparing two ligands, the larger one is denoted ligand L ($\thetL$), and the smaller one is denoted ligand S ($\thetS$). Similarly, we refer to the cognate ligand ($\thetC$, \ie~the target), and the non-cognate ligand(s) ($\thetNC$, the functionally undesirable one).
  
  For each protein-ligand pair, we calculate the free energy of binding $\DG$ (\ie, the binding affinity) as the sum over three contributions: the deformation energy $\edf$, the chemical energy $\ech$, and change in entropy upon binding $\DS$,
  \begin{equation}\label{eq:DG}
  \DG = \edf + \ech - T \DS~,
  \end{equation}
  where $T$ is temperature. Deformation energy is calculated as
  \begin{equation} \label{eq:edef}
    \edf = \frac{1}{2} \sum_{\langle i, j \rangle}  K_{ij} (\Dr_{ij} - \ell_{ij})^2~,
  \end{equation}
  where the summation $\langle i, j \rangle$ is over all ordered pairs of amino acids $a_i$ and $a_j$ connected by a bond, $K_{ij}$ is the spring constant of the bond, $\Dr_{ij}$ is its length, and $\ell_{ij}$ its equilibrium length. The initial configuration is not deformed, $\edf=0$, since all distances are equal to the equilibrium bond lengths, $\Dr_{ij} = \ell_{ij}$. In the \textbf{MeCh} model all bonds have identical lengths, $\ell_{ij} = \ell = \SI{1}{nm}$, and spring constants, $K_{ij} = K$.
  
  The chemical binding energy is given by
  \begin{equation}\label{eq:ech}
      \ech = \sum^{\NB}_{i=1} -\epsilon_i \, 
      e^{-|\rb^b_i - \rb^p_i|^2/\sigma^2}~,    
  \end{equation}
  where $\epsilon_i$ is the energy scale of binding locus $i$ of the ligand, $\rb^p_i$ and $\rb^b_i$ are the positions, respectively, of the amino acids $a_j$ at the binding pocket and the ligand binding locus $i$, and we set the length scale of the interaction at $\sigma = \SI{0.3}{\nm}$; interaction sites are uniquely paired, so the energy is summed only over the $\NB=3$ binding pairs. In the \textbf{MeCh} model the energy scale is the same for all binding sites, $\epsilon_i = \epsilon$.
  
  The entropy change by binding $\DS$ in \eref{eq:DG} is the logarithm of the relative change in the volume of the configuration space accessible by thermal fluctuations of the protein. Thus, reduction in the magnitude of thermal motion results in entropy loss.  
  We decompose the binding entropy into two terms, $\DS = \DSC + \DSO$, where $\DSC$ is the change in conformational entropy,~\cite{frena07,tzena12,sunjc17} which depends on the elastic network topology and spring stiffness, and $\DSO$ accounts for other contributions to entropy that are not captured directly by the model (\eg, release of frustrated solvent, ligand conformational change, protein entropy change due to plastic deformation).~\cite{fenpn12,draeb17,quina89,micja09,pecao21,keuna18,wanel22}

   We calculate $\DSC$ for the elastic network by creating stiff bonds of strength $\KL$ between the protein and ligand binding sites (see Methods). Standard normal mode analysis shows that the resulting entropy change is the sum over the variation in the logarithms of the mode energies $\lambda_n$ before and after binding, $\DSC =
  -\half \sum_n {\Delta\ln\lambda_n}$. By constraining the motion, stiffening typically increases the mode energies, $\Delta\lambda_n \ge 0$, and binding therefore induces entropy loss, $\DSC \le 0$.
  For a homogeneous spring network ($K_{ij} = K$), this entropy change is well approximated as $\DSC \approx -2\ln(K / \KL) + 1.5$ (SI Fig. 1).
  
  We compute the minimum binding free energy using a gradient descent algorithm with some relevant physical constraints. Calculating this binding free energy for multiple ligands gives us the binding free energy gap between any pair of two ligands, $L$ and $S$, $\DDGLS = \DGL - \DGS$ (or between cognate and non-cognate ligands, $\DDG=\DGNC - \DGC$), which we use as a measure of molecular discrimination and specificity.\\

\nsb{Recognition via lock-and-key binding} 
 As a starting point, we examine the limiting case of a completely rigid protein (\ie, $K=\infty$), which corresponds to \textit{lock-and-key} binding.~\cite{fisbd94}
 As no deformation occurs upon binding and no conformational entropy is lost (if we set $\DSO=0$), the binding free energy only depends on the mismatch $\dlp = \theta - \theta_0$ of the shape of the wedge and the ligand:
  \be
  \DG_{\rm rigid} = \ech(r) = 
  \begin{cases}
    -\epsilon,& \text{if } \theta > \theta_0\\
    - \epsilon \l( 2 +  e^{-r^2 / \sigma^2} \r) ,& \text{if } \theta \leq \theta_0\\
  \end{cases}~,
  \ee
  where $r = 2\ell\sin{\half\dlp} \approx \ell \dlp$ is the gap between the third interacting pair of the binding site and the ligand. This is because fat ligands can only interact via one site, and thin ligands minimize binding energy by fully binding to two sites, and partially binding to the third.

  In this limiting case, the best binding gap, $\DDG = 2\epsilon$, is
  achieved via \textit{steric exclusion}: when one ligand is a perfect
  match for the binding site ($\theta_0 - \thetS = 0$), and the other
  ligand is fat, for any $\dlig = \thetL - \thetS$
  (\fref{fig:fig1}B). If both ligands are thin, the largest binding gap is much lower  ($\DDG=\epsilon$),~\cite{tawfj20} and can only be obtained for sufficiently dissimilar ligands ($\dlig \geq \ang{40}$). If the two thin ligands are similar, the optimal binding gap is only achievable with some \textit{mismatch} between the cognate ligand and the protein ($\thetL < \theta_0$). We thus see that binding with a rigid protein is in principle a feasible strategy for molecular discrimination. However, even in this hypothetical case, shape mismatch is often necessary to promote binding of the cognate ligand while avoiding binding of the non-cognate.
\\

  \begin{figure*}[t!]  \centering
  \includegraphics[width=0.99\linewidth]{fig2_2.pdf}
  \caption{\label{fig:fig2}
  \nsb{Effect of energy and entropy on binding and discrimination in the \textbf{MeCh} model}
  \sb{A}: Binding free energy for ligand L vs. ligand S; $\epsilon=\SI{8}{kT}$, $-T\DSO=\SI{10}{kT}$, $\KL=\SI{e4}{kT/nm}^2$. Colored regions indicate ``loose discrimination'' ($\DGC < 0$ and $\DDG \geq \SI{2}{kT}$), or ``strict discrimination'' ($\DGC < 0$, $\DGNC > 0$ and $\DDG \geq \SI{2}{kT}$).
  Binding free energy is shown for the proteins from Fig.~\ref{fig:fig1}C-D ($\dlig=\ang{5}$).
  \sb{B}: Binding free energy gap as a function of spring constant $K$, and chemical energy constant $\epsilon$; $\KL=\SI{e4}{kT/nm}^2$, $\thetC=\ang{95}$, $\thetNC=\ang{100}$. Optimal $K$ for a given $\epsilon$ is indicated by the orange dashed line.
  \sb{C}: Regions that satisfy the conditions for loose and strict discrimination for different values of $-T\DSO$.
  \sb{D}: Affinity ($\DG$) versus specificity ($\DDG$) for the parameter space sampled in (B-C).
  }
  \end{figure*}

  \nsb{Preferential binding by flexible proteins exploits ligand-protein shape mismatch}
  Rigidity in proteins is limited by the nature of the non-covalent bonds holding them
  together. To explore the effect of rigidity on discrimination, we first set wide bounds on
  the spring constants ($10 \le K \leq \SI{e4}{kT/\nm^2}$), and later narrow these to
  more biologically-plausible values of spring constants.~\cite{atibj01,hinbi07,ricbj09}
  When $K$ is low, binding is not very specific as the protein can perfectly match any ligand with little deformation energy (\fref{fig:fig1}C). As $K$ increases, specificity improves, until the protein is so rigid that neither cognate nor non-cognate ligands can induce the protein to deform. The maximum binding free energy gap for fat ligands ($\DDG = 0.84 \epsilon$, $\dlig=\ang{5}$, $K= \SI{e4}{kT/\nm^2}$) is considerably lower than what can be achieved by rigid lock-and-key binding ($\DDG = 2 \epsilon$). 
  
  For thin ligands, on the other hand, the maximum gap 
  ($\DDG = 0.49 \epsilon$, $\dlig=\ang{5}$, $K\sim \SI{3e3}{kT /\nm^2}$) is double (\fref{fig:fig1}C) of what can be achieved by a completely rigid protein ($\DDG=0.24 \epsilon$, $\dlig=\ang{5}$). In general, higher specificity is still achieved with greater rigidity across biologically-relevant values of $K$;   but this requires ever-greater shape match between the protein and the cognate ligand, as $\DDG$ becomes a steeper function of ligand shape (\fref{fig:fig1}D). This coupling between flexibility and shape mismatch   highlights the need for precise concerted control over both protein structure and dynamics.
\\

  \nsb{Discrimination is more difficult than recognition}
  The binding energy gap is a key determinant of molecular discrimination, but binding will only occur spontaneously if the corresponding free energy change $\DG$ is negative, which depends on binding entropy.~\cite{frena07,tzena12} 
  Taking this into account, we now formally define molecular discrimination:   `loose' discrimination is defined by $\DDG \geq \SI{2}{kT}$, and $\DGC < \SI{0}{kT}$; we use \SI{2}{kT} as a reasonable threshold that corresponds to a sevenfold difference in binding affinity; `strict' discrimination is defined by further requiring $\DGNC > \SI{0}{kT}$.
  
  In \fref{fig:fig2}A we replot the data from \fref{fig:fig1}C-D in terms of discrimination, finding that specific binding is substantially more difficult than recognition alone. Even for our permissive threshold of \SI{2}{kT}, only a fraction of cases result in discrimination. The disparity between the ability to distinguish between sets of fat ($\DGS < \DGL$; upper left) versus thin ($\DGL < \DGS$; bottom right) ligands
  illustrates the utility of steric exclusion as a discrimination mechanism (discrimination is possible over a large range of $\DSO$ for fat ligands, but if the ligands are thin then it is difficult to achieve discrimination). \\

  \nsb{Affinity and specificity are correlated}
  To understand the role of mechanical and chemical binding energy in discrimination, we now vary the chemical energy constant so that the chemical binding energy is within a reasonable biological range ($1 \leq \epsilon \leq \SI{25}{kT}$) .~\cite{schjm99,barja05,galja10,moabi11,huach13} Once again, increasing $K$ aids specificity, up until a point where the protein is too rigid to accommodate either ligand (\fref{fig:fig2}B).  Beyond this point, increases in $K$ must also be accompanied by higher $\epsilon$, so that there is sufficient chemical driving force for mechanical deformation. $\epsilon$ sets an upper bound on specificity ($\DDG$), which is achieved at some optimal $K$, and this optimal $K$ is proportional to $\epsilon$. As a result, affinity ($\DGC$) and specificity ($\DDG$) are naturally correlated within the parameter space (\fref{fig:fig2}D, Pearson's $r=0.38$, $p\ll0.05$).\\

  \nsb{Entropy can be used to modulate affinity}
  Binding entropy is a key determinant of molecular discrimination   (\fref{fig:fig2}C).~\cite{barja05,grust06,chapn07,huach13,sunjc17} For the same discrimination task as in \fref{fig:fig2}B, only loose discrimination is possible for $-T\DSO=\SI{0}{kT}$ since both ligands bind to the protein; for this particular case, in the regions with a large binding energy gap, $\num{3} < -T\DSC < \SI{10}{kT}$, is too low to avoid binding to the non-cognate ligand. As we increase the entropy cost $-T\DSO$, loose discrimination becomes more difficult since stronger chemical bonds are needed for binding. However, increasing $-T\DSO$ also enables strict discrimination, eventually to the point where the entropic cost is too high for any ligand to bind.
  Since we have assumed that entropic cost applies equally to cognate and non-cognate ligands, it does not affect
  $\DDG$; tuning entropy thus offers a way to decouple affinity and specificity (\fref{fig:fig2}D).\\

  \begin{figure*}[htb!]  \centering
  \includegraphics[width=0.99\linewidth]{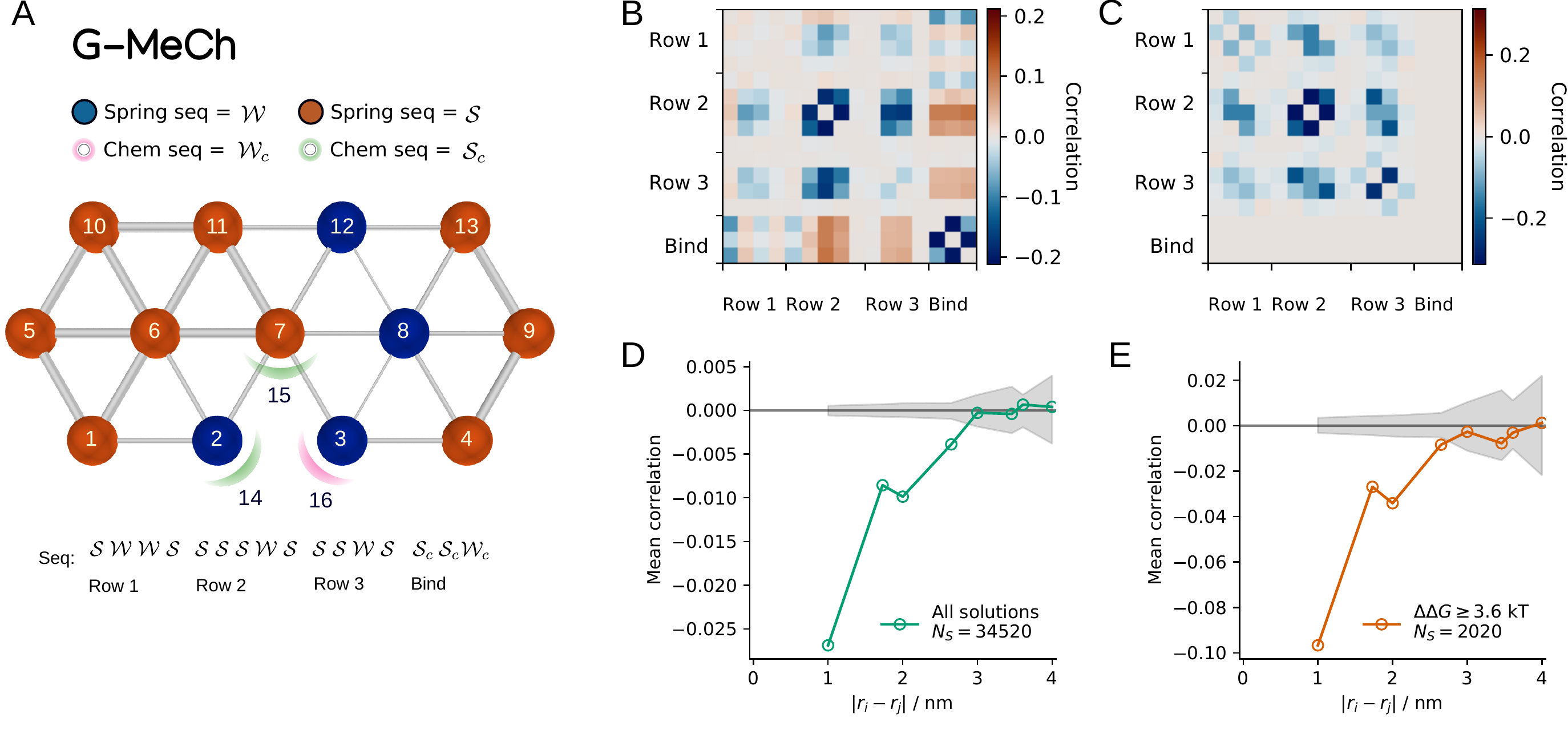}
  \caption{\label{fig:fig3}
  \nsb{Linking mechanics and binding chemistry to genetic sequence in the \textbf{G-MeCh} model}
  \sb{A}: Amino acids are either $\LW$ (blue, forming weak bonds), or $\LS$ (orange, forming strong bonds), as indicated by the colour of the spheres; edge width indicates the three possible values of bond strength of the bonds as determined by the sequence; chemical energy is similarly determined by sequence as $\LLW$, or $\LLS$ as indicated by the colour of the haloes at the interaction sites. Sequence is read from left-to-right, from bottom-to-top (as indicated by the numbers).
  \sb{B-C}: Sequence correlation for all loose-discrimination solutions (B), and solutions with $\DDG > 3.6$ kT (C).
  \sb{D-E}: Mean sequence correlation versus distance between residues; shaded region indicates the standard deviation from the mean sequence correlation from $N_S$ random sequences.
  For this example, the parameters are: $\Ks=\num{e3}$, $\Km=\num{400}$, $\Kw=\num{100}$, $\KL=\SI{e4}{kT/\nm^2}$; $\epss = 6$, $\epsw = \SI{3.88}{kT}$; $\thetS=\ang{80}$, $\thetL=\ang{90}$; $T\DSO=\SI{0}{kT}$.
  }
  \end{figure*}

  \nsb{Sequence-dependent model (G-MeCh)}
  We have so far ignored the quintessential feature of proteins -- proteins are heteropolymers, composed of distinct amino acids whose sequence is encoded in genes that are subject to evolution. We now expand our model to include these two important components -- heterogeneity in flexibility, and change via discrete mutations -- by coupling the mechanics and chemistry to the protein sequence. 
  
  Thus, we examine a model protein consisting of $\NA=13$ amino acid letters, and $\NB=3$ binding letters, using a 2-letter alphabet for each (\fref{fig:fig3}A). Amino acids are either $\LW$ or $\LS$, and the elastic bond between each pair of neighbours, $a_i,a_j$, depends on their identities $K_{ij}(a_i,a_j)$: bonds are either strong ($\LS{-}\LS$, with spring constant $\Ks$), medium ($\LS{-}\LW$ or $\LW{-}\LS$, $\Km$), or weak ($\LW{-}\LW$, $\Kw$). Similarly, chemical binding strength is determined by the letters $\LLW$ and $\LLS$, respectively, resulting in weak ($\epsw$) and strong bonds ($\epss$). The change in conformational entropy is typically greater for more flexible proteins, and is incorporated in the model by comparing the conformational entropy of the spring network before and after binding (Methods).\\

  \nsb{Sequence variation reveals epistasis}
  For a representative set of parameters, we calculate the binding free energy $\DG$ for all possible $2^{\NA +\NB}=2^{16}=\num{65536}$ sequences, and identify those sequences that result in discrimination. On average, there is no bias towards a particular amino acid at any position (SI Fig. 2), with the exception of strong chemical binding at the tip of the $\Lambdaup$-shaped binding pocket (position 15). 
  However, detailed examination of sequence correlation (Methods) reveals substantial, non-random patterns of \textit{epistasis} (\fref{fig:fig3}B). There are positive correlations between the chemical binding sites and the central amino acids in rows 2 and 3 -- increasing the stiffness of the intramolecular bonds made by these amino acids allows for stronger chemical binding. One could anticipate this, given the correlations between $\epsilon$, $K$ and $\DDG$ found in \fref{fig:fig2}B. Thus, the subset of solutions with $\DDG>\SI{3.6}{kT}$ (\fref{fig:fig3}C) contains only those sequences that have taken advantage of this positive correlation -- all sequences have strong chemical bonds, and the central amino acids typically have strong bonds (SI Fig. 2).\\

  \nsb{Far-away residues enable fine-tuning of protein binding}
  All epistasis visible in \fref{fig:fig3}C is the outcome of negative correlations in the sequence. For two positions that negatively covary, when one position goes from weak to strong (or vice versa), the other position tends to do the opposite.
  In this way protein flexibility is fine-tuned, to achieve a nearly-optimal amount of deformation at the binding site. This is most apparent in the correlation between all of the central seven amino acids. This can be explained by the existence of many different sequences that can encode the same open-close motion at the binding site (SI Fig. 3). Note, however, that this effective fine-tuning of the protein mechanics is not limited to the binding site and is evident throughout the full length of the protein, between pairs of residues at distances up to \SI{2.5}{\nm} (\fref{fig:fig3}D-E). \\

  \begin{figure}[t!]  \centering
  \includegraphics[width=0.99\linewidth]{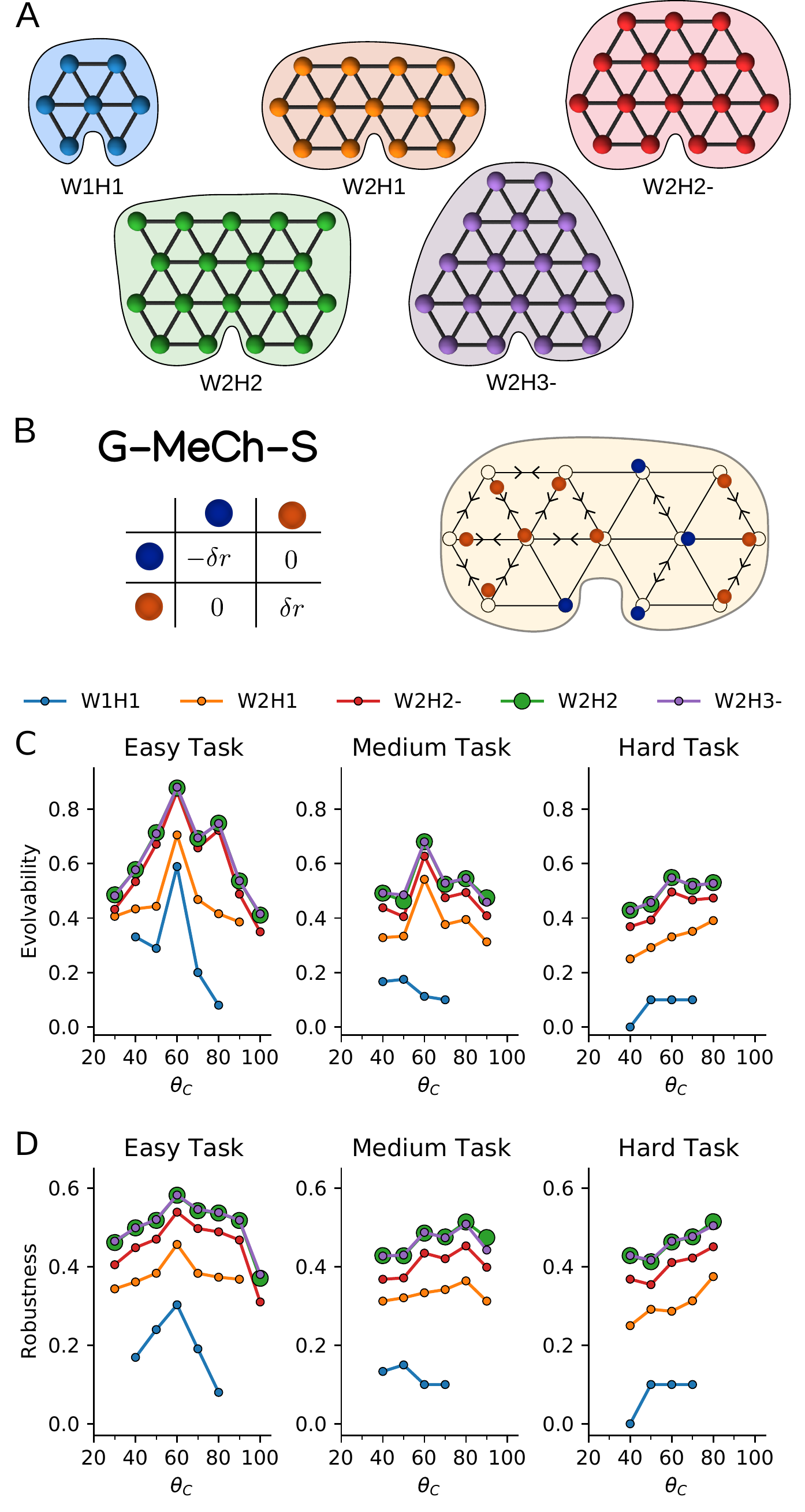}
  \caption{\label{fig:fig4}
  \nsb{The G-MeCh-S model}
  \sb{A}: Protein models of different size.
  \sb{B}: Protein sequence is mapped to perturbations in the protein structure according to an interaction matrix. If the interaction is positive, neighbouring amino acids experience an attractive force with displacement magnitude $\delta r = \SI{0.1}{\nm}$ and vice versa (black arrows); the overall displacement is the average of the interactions with all neighbouring amino acids. The open black circles are the original positions, and the coloured circles show both the identities of the amino acids and the new equilibrium configuration.
  \sb{C-D}: Evolvability (C) and robustness (D) (both normalized by sequence length) as a function of the angle of the cognate ligand, $\thetC$, for easy, medium, and hard discrimination tasks. Colours indicate protein models of different size.
  Green circles are large to highlight that they overlap with purple circles.
  Symbols are only shown for ligands for which solutions were found. Parameters are: $\Ks=\num{e3}$, $\Km=\num{400}$, $\Kw=\num{100}$, $\KL=\SI{e4}{kT/\nm^2}$; $\epss = 6$, $\epsw = \SI{3.88}{kT}$; $T\DSO=\SI{0}{kT}$.
  }
  \end{figure}

  \nsb{Structural perturbation model (G-MeCh-S)}
  Residues far from the binding site give the protein extra degrees of freedom that can be used to fine-tune binding. We thus study how these degrees of freedom are related to the ability to evolve functional sequences that are robust to mutations, by modelling proteins of different size (\fref{fig:fig4}A).
  To study protein evolution, we map genotype to fitness via a simple measure $F$ of binding specificity,\cite{savce13}
  \begin{equation}\label{eq:fit}
    F = \frac{e^{-\DGC} - \sum_{i} {e^{-\DGNC^i}} }{1 + e^{-\DGC} + \sum_{i} {e^{-\DGNC^i}}}~,
  \end{equation}
  where $\DGC$ is the binding energy of the cognate ligand, and the non-cognate ligands have binding energies $\DGNC^i$ (all energies are in kT units). The performance measure, $-1 \le F \le 1$, is the difference between the binding probability of the cognate ligand and the sum of probabilities to bind any other competitor, and sequences are considered fit if $F > 0$.

The structure of real proteins varies with their sequence, generating elaborately rugged fitness landscapes, with numerous maxima and basins of attraction.~\cite{hiepn11,devnr14} While the \textbf{G-MeCh} model generates many fitness maxima, they are rather shallow, forming large basins of nearly-optimal configurations. This is because the shape of the \textbf{G-MeCh} protein is sequence-independent, rendering the fitness landscape unrealistically smooth. 
Therefore, we introduce the \textbf{G-MeCh-S} model, where the protein \textit{equilibrium} structure is allowed to change depending on sequence variation -- a more realistic approximation.
The resulting affinity, specificity, and fitness landscapes become much more rugged, containing multiple basins of attraction (SI Fig. 4). As a protein's shape varies, it evolves to switch preferences among competing ligands. Note that we do not model protein folding, but instead account for sequence variation resulting in small perturbations to the native structure (\fref{fig:fig4}B, Methods).

  We study the effect of protein size on the evolution of proteins that can discriminate a target ligand from many non-cognate ligands. We successively increase task difficulty by increasing the number of non-cognate ligands. For each task, the target ligand is picked from a set of ligands, $\theta \in \{\ang{20}, \ang{30}, \ldots, 100^{\circ}\}$. For each cognate ligand $\thetC$
  we designate its neighbours as competing non-cognate ligands  (\ie, those which satisfy $|\thetC - \thetNC| \leq \phi$). This allows us to consider discrimination tasks of variable difficulty:
  easy, $\phi = \ang{10}$, medium, $\phi = \ang{20}$, and hard,  $\phi = \ang{30}$, such that in the easiest task there are at most 2 non-cognate ligands, and in the hardest task there are at most 6 non-cognate ligands. In this way, when the target ligand has high shape mismatch, there are fewer non-cognate ligands, so task difficulty is not strongly dependent on $\thetC$. In principle, one could have also varied task difficulty by changing $\dlig$, but the present approach has practical advantages (SI Fig. 4).

  \nsb{Larger proteins are more evolvable and robust}
  We quantify evolvability as the maximum number of mutations a fit protein can undergo while still remaining fit ($F\geq0$). Conversely, we quantify robustness as the minimum number of mutations a fit protein can undergo before it becomes unfit.
  We find that larger proteins are able to solve discrimination tasks for a greater range of cognate ligands, and longer sequences
  are more evolvable and robust,~\cite{levns09,tawar10,wag13}
  even after controlling for sequence length (\fref{fig:fig4}C-D, Methods).
  Increasing protein size gives diminishing returns which eventually saturate, indicating that finite degrees of freedom are sufficient to achieve a maximally evolvable and robust protein.
  For more difficult tasks, proteins of different size exhibit a
  larger difference in evolvability and robustness, and larger proteins are needed to reach the saturation point (\fref{fig:fig4}C-D, SI Fig. 5). Thus, the required degrees of freedom depend on the difficulty of the discrimination problem.\\

  \begin{figure*}  \centering
  \includegraphics[width=0.99\linewidth]{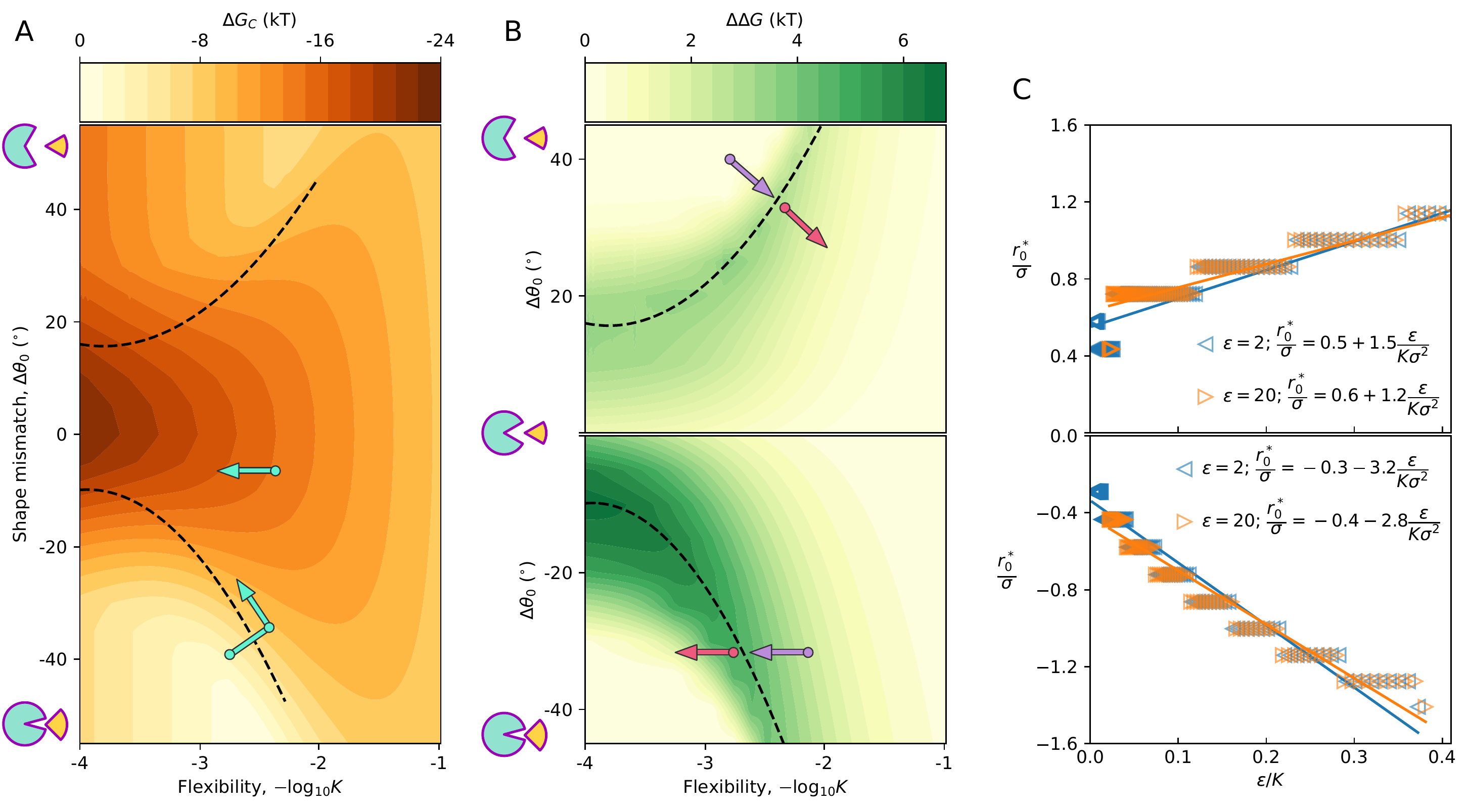}
  \caption{\label{fig:fig5} 
  \nsb{Phase diagram of affinity and specificity for the \textbf{MeCh} model}
  A-B: Binding affinity (A) and specificity (B) as a function of protein
  flexibility, and shape mismatch between cognate ligand and binding pocket
  ($\epsilon=\SI{8}{kT}$, $|\dlig|=\ang{10}$, $-T\DSO=0$, $\KL=\SI{e4}{kT/\nm^2}$).
  In the top plot (B) the cognate ligand is smaller than
  the non-cognate, and the binding pocket is open to the ligand; in the bottom plot (B) the cognate
  ligand is larger than the non-cognate, and the binding pocket is narrow compared to the ligand.
  The optimal mismatch for a given flexibility is shown by the black line.
  Cyan arrows (A) indicate proposed experimental trajectories of antibodies during affinity maturation.~\cite{ovcel18}
  Purple arrows (B) indicate proposed experimental trajectories of enzyme evolution (top -- esterase,
  bottom -- cytochrome P450), and red arrows indicate predicted trajectories of suggested experiments.~\cite{marac18,nutjc21}
  C: Scaling of optimal mismatch $r^*$ with chemical energy $\epsilon$ and stiffness $K$.
  Fits (lines) to model results (symbols) are shown for open (top) and closed (bottom) binding pockets, and for $\epsilon=2$ and $20$ kT. Symbols tend to overlap.
  }
  \end{figure*}

  \nsb{Tuning affinity and specificity via the interplay of energy, entropy, and  shape} 
  Proteins have an optimal affinity range for cognate ligand(s),~\cite{popbi20} and require
  some degree of specificity for functionality. The \textbf{MeCh} model explains how these can be controlled by varying protein flexibility, shape, chemistry, and entropy;
  this is illustrated in a phase diagram of affinity and specificity (\fref{fig:fig5}A-B).
  We study a smaller parameter space for the \textbf{G-MeCh} model, and find qualitatively similar results (SI Fig. 6). 
  
  Affinity ($\DG$) depends non-linearly on chemical bond strength, shape mismatch,
  pocket geometry (open / closed), flexibility, and entropy (\fref{fig:fig5}A).
  The number, and strength of potential chemical bonds (represented in the model by $\epsilon$)
  sets an upper limit to affinity (if $-\DSO \geq 0$; see ``Limitations'' section for counter-examples not covered here),
  which is attained when a ligand fits perfectly into a rigid protein pocket.
  For rigid proteins, shape mismatch sharply reduces affinity, but still results in
  zero-deformation partial binding, which is stronger in proteins with open binding pockets.
  Flexibility can both increase and decrease affinity:
  flexible proteins have higher conformational entropy, which may result in higher binding entropy cost and weaker affinity;~\cite{caops16}
  but if there is some shape mismatch between the ligand and the binding pocket,
  flexibility increases affinity by enabling the protein to adopt an optimal conformation for binding, as clearly demonstrated by our model (\fref{fig:fig5}A).
  
  The specificity $\DDG$ for a pair of cognate and non-cognate ligands differing by $\dlig = \ang{5}$ is depicted in \fref{fig:fig5}B. 
  Maximum specificity for a given $\epsilon$ is achieved by matching flexibility and shape mismatch (\fref{fig:fig5}B),
  so that deformation energy is maximized (SI Fig. 7).~\cite{ricja19} 
  Proteins need to be just flexible enough for the cognate ligand to bind, but not the non-cognate ligand.
  Too much flexibility (reduces deformation energy) and both will bind;~\cite{gadja21}
  too rigid and both will partially bind with negligible deformation. The optimal deformation
  for a given flexibility is found to scale linearly with the chemical binding energy $\epsilon$ (\fref{fig:fig5}C), such that the overall shape of the specificity phase diagram is not affected by changes to $\epsilon$ (SI Fig. 8). We will later demonstrate how these findings can be recapitulated by a simplified phenomenological model.

  The model predicts how entropy may affect affinity and specificity:
  The primary effect of entropy in the \textbf{MeCh}-model is to modulate affinity via changes in conformational entropy $\DSC$ as the flexibility varies (\fref{fig:fig5}A). This stiffening effect will be similar for competing ligands and thereby will impact specificity only weakly (although for sufficiently dissimilar ligands, this contribution of entropy to specificity may be more significant). Thus, by selectively tuning entropy, one can modulate affinity almost independently of specificity (\fref{fig:fig2}D). In practice, this might be difficult; for example, flexibility simultaneously affects binding energy and conformational entropy.~\cite{grust06}
  There is more potential for independent control via the other components to binding entropy, $\DSO$. Examples of this include the use of \textit{intrinsically-disordered} regions:
  through coupled folding and binding;~\cite{wanpn13_2,yanps19} or through
  a disordered region far from the binding site, which is affected allosterically.~\cite{keuna18}
  Other routes to decoupling specificity and affinity include control over
  oligomeric complexation,~\cite{schpn21} or interactions with solvent.~\cite{quina89,micja09} \\

 \nsb{Phenomenological model} The general nature of the
    specificity can be illustrated by a simple phenomenological model (Methods): the free energy is taken as a function of a single coordinate $r$, which represents the gap between a pair of interacting loci of the binding site and the ligand. The initial gap $r_0$, before the protein deforms, is proportional to the mismatch, $r_0 \approx \ell \dlp$. The elastic energy is a function of the deformation $u  = r_0 - r$, while the chemical energy is a function of the gap $r$, to which we add binding entropy to obtain the free energy (\eref{eq:DG}), $\DG(r,r_0) = \edf(u) + \ech(r) - T \DS~$.
    This protein-ligand system reaches an equilibrium configuration, $r=\bar{r}$ when the elastic and chemical forces counterbalance each other $-\edf'(u) + \ech'(r) = 0$ (where $'$ is the derivative). 
    
    The specificity $\DDG$ in discriminating between a cognate ligand and a competitor, whose initial gaps differ by $\delta r_0$, is the change in $\DG$ along the equilibrium line, $\bar{r}(r_0)$.
    Analysis shows that the specificity is proportional to the forces,  $\DD = \DDG/\delta r_0 \approx \edf'(u) = \ech'(r)$ (where $\DD$ is the ``specific specificity'', or specificity per unit of difference between ligands).
    Therefore, maximal specificity is obtained exactly for the mismatch $r_0^\ast$ and gap $r^\ast$ at which the opposing elastic and chemical forces are both maximal, at the inflection point,
    $\ech''(r) = 0$, where the system is most sensitive to small shape differences (see details in Methods). This simple result explains why the line of optimal specificity follows the line of maximal deformation energy in our model (SI Fig. 7).

    We apply this general theory to \textbf{MeCh} model, and consider the standard
    Hookean elastic energy $\edf(u) = \half \Keff u^2$
    (\eref{eq:edef}). The whole protein is expected to be softer than
    a single bond, with an effective spring constant $\Keff = \alpha
    K$ ($\alpha \le 1$). As for the chemical energy (\eref{eq:ech}),
    we assume that two loci of the $\Lambdaup$-shaped ligand are in
    contact with the (open) binding pocket and contribute $-2\epsilon$ to the
    free energy, which together with the energy of the third locus
    gives $\ech(r) = -\epsilon (2 + e^{-r^2/\sigma^2})$. The
    inflection point of this chemical energy is located at $r^\ast =
    \sigma/\sqrt{2}$, such that the optimal mismatch is  
    \begin{equation}\label{eq:optmis}
    \frac{r_0^\ast}{\sigma} =  2^{-\half}+ (2/e)^\half \l( \frac{\epsilon}{\Keff \sigma^2} \r)
    \approx 0.71 + \frac{0.85}{\alpha} \l(\frac{\epsilon}{K \sigma^2}\r)~,
    \end{equation}
    with a corresponding optimal ``specific specificity'',
    $\DD^\ast= \DDG^\ast/\delta r_0 \approx 0.85 (\epsilon/\sigma)$.
    
    We fit \eref{eq:optmis} to the \textbf{MeCh} model predictions for the optimal mismatch (\fref{fig:fig5}C). For the case of an open binding pocket ($\dlp > 0$), we find an intercept of $0.5-0.6$ and effective softening $0.57 < \alpha < 0.71$, in broad agreement with the phenomenological model. For a narrow pocket, we find qualitatively similar behaviour: due to steric exclusion, less mismatch is needed for a maximally rigid protein (\ie, the intercept is closer to zero); the slope of the $r_0^\ast/\sigma$ line is approximately double in this case, since the reduction in affinity due to steric exclusion is roughly twofold ($2\epsilon$) compared to the open pocket case (\fref{fig:fig1}B). 
    The phenomenological model (\eref{eq:optmis}) elucidates two findings of the detailed protein models: (i) Mismatch is needed even for extremely rigid proteins (\fref{fig:fig5}B). (ii) The optimal mismatch depends on a trade-off between chemical binding energy and mechanical energy, as demonstrated in \fref{fig:fig2}B where optimal specificity is achieved at a constant ratio  $\epsilon/K$. In principle, this simple fit will also apply to the \textbf{G-MeCh} model, but there $\Keff(\{ K_{ij}\})$ has to be extracted for each protein (SI Fig. 6).

\section*{\sb{Discussion}}
  Through analysis of a conceptually simple, yet multi-faceted model of molecular discrimination, we start to understand the mechanisms, and evolution, of molecular discrimination. \\

  \noindent\sb{How difficult is molecular discrimination?~}
  It depends on the context. For example, proteins that bind to nucleic acids or lipids must find their target out of a dizzying gallery of lookalikes.~\cite{kriar19,jannr15,conna12} Enzymes need to be able to release their product after catalysis,~\cite{olijr20} which can be more (\eg, isomerization reactions) or less (\eg, proteolysis) challenging depending on how similar the product is to the substrate.
  In principle, the difficulty, $\Psi$, can be expressed as a function of the set of ligands, $\{ \Lambdaup \}$, and the required degree of specificity, 
  \begin{equation*}
  \Psi \left [ \Lambdaup^\Cs, \DGC, \Lambdaup^\NCs_1, \DDG_1, \ldots~,  \Lambdaup^\NCs_i, \DDG_i \right ]~,
  \end{equation*}
  where $\DGC$ is the optimum binding affinity of the cognate ligand, and $\DDG_i$ is the required minimal binding energy gap between the cognate ligand $\Lambdaup^\Cs$ and non-cognate ligand $i$, $\Lambdaup^\NCs_i$; the degree of specificity required for each   non-cognate ligand depends on the \textit{in vivo} concentration,~\cite{gadja21} and the cost of incorrect binding.
  
\begin{figure}  \centering
  \includegraphics[width=0.99\linewidth]{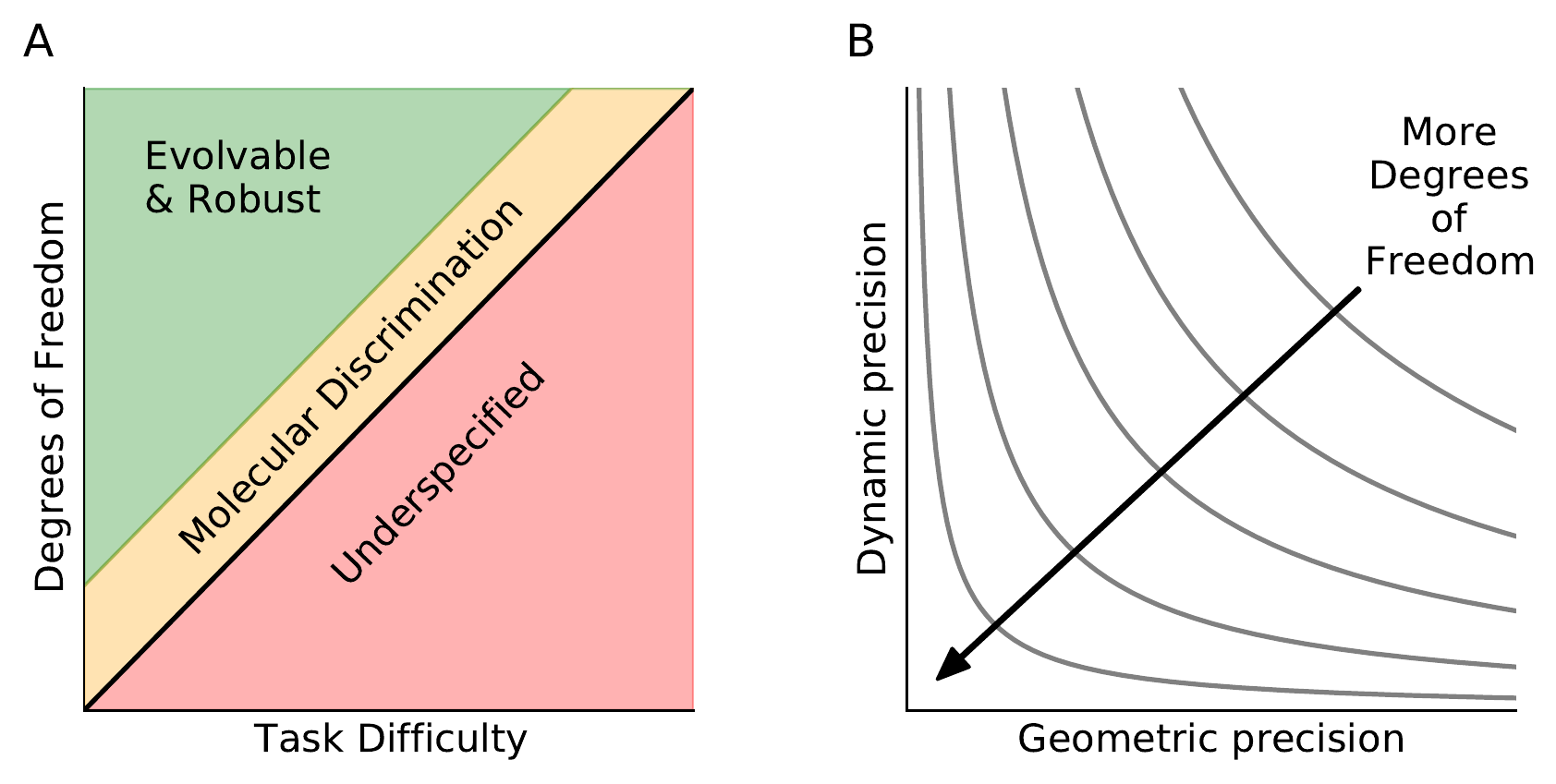}
  \caption{\label{fig:fig6} 
  \nsb{Theory of biomolecular discrimination}
  \sb{A}: More difficult discrimination tasks require more degrees of freedom,
  which in proteins corresponds to longer sequences. Sequences that are
  longer than the minimum necessary size are more robust and evolvable.
  \sb{B}: Larger proteins have more degrees of freedom with which to fine-tune
  structure and dynamics at the binding site, since mutations at distal
  residues can have small yet significant effects.
  }
  \end{figure}

  The difficulty $\Psi$ is a non-linear function, which we expect to generally increase with: the required specificity ($\DDG_i$), the number of similar, non-cognate ligands $\Lambdaup^\NCs_i$, and how easy it is to distinguish them from the cognate ligand $\Lambdaup^\Cs$. It is easier to distinguish: if the cognate ligand is smaller than the non-cognate, as it allows steric exclusion of the larger ligand (\fref{fig:fig1}B); if the cognate ligand can form more energetic bonds (\eg, an extra hydroxyl group effectively increases $\epsilon$);~\cite{pertb18} or has distinct chemical differences   (\eg, positive versus negative charge) to the non-cognate ligands.~\cite{barja15} We lack a general, robust method of quantifying $\Psi$; creating a metric of ligand discriminability (not the same as similarity~\cite{stejm13,racjc18}) would be immensely useful for, \eg, predicting  sites for specific inhibition of proteins. For now, we propose that an ad hoc approximation for $\Psi$ could be taken to be the degrees of freedom needed to achieve functional discrimination (\fref{fig:fig6}A).\\


  \nsb{Molecular discrimination by a hypothetical mechano-chemical machine}
  Lock-and-key binding is most specific, but only under very strict conditions (\fref{fig:fig1}B): The non-cognate ligands must be larger than the cognate ligand (steric exclusion), the cognate ligand must perfectly match the binding site, and both protein and ligand must be rigid. Otherwise, some mismatch between cognate ligand and binding site is always needed to optimise specificity (\fref{fig:fig1}D)  -- such `conformational proofreading' (\ie, an energetic penalty) minimises binding to the non-cognate ligand(s).~\cite{savpo07,savce13,savmc10,devmc12} The path to better discrimination still lies with higher rigidity, but increasing control over shape mismatch is needed, since deviations from the optimal mismatch are tolerated less and less as rigidity increases (\fref{fig:fig1}D). Ultimately, the key feature of good discrimination is precision: the right amount of flexibility, coupled with the right shape, results in optimal deformation.\\

  \nsb{Molecular discrimination by proteins}
  Proteins, as \emph{genetic}-mechano-chemical machines, have some inherent features that strongly constrain their molecular discrimination ability.~\cite{pertb18,sikjr14} Most interactions between amino acids are non-covalent, so proteins cannot be very rigid.   Moreover, protein geometry is limited to the topology of a folding chain, composed of discrete units (amino acids) of approximately \SI{1}{\nm} in size, so perfect shape match is practically impossible. Finally, evolution advances in discrete steps, mutation, deletions and insertions -- not via continuous tuning.
  
  In light of these constraints, mutations very close to a binding site are bound to have a large effect on the flexibility and geometry of the binding site.~\cite{lopjc10} On the other hand, mutations further from the binding site can have ever smaller effects,~\cite{adacs19,wannm20} enabling fine-tuning of mechanics (\fref{fig:fig3}),~\cite{koepn17,ottnc18,modnc21,osepc22,raypn21,chrpn22} and structure.~\cite{marsc21,adics21} 
  For example, in a separate recent study of ours, we studied the effect of single mutations on structure (using proteins in the protein data bank, and AlphaFold), finding that structural perturbations are felt up to \SI{2}{nm} away from the mutated residue.~\cite{mcbbi22}
  This is consistent with our assertion that far-away mutations can influence molecular discrimination.
  
  These observations, taken together with the finding that good discrimination necessitates precision, lead us to propose that larger proteins -- as they have more degrees of freedom (potential mutations) -- can achieve better discrimination through finer control over protein dynamics and structure at the binding site (\fref{fig:fig6}B). For a given discrimination task there is a minimum protein size (\fref{fig:fig4}C-D), but such an efficient protein may be difficult to evolve. As protein size grows beyond the bare minimum, there are ever more sequences capable of solving the problem, which results in sequences that are evolvable and robust (\fref{fig:fig4}C-D). The proposed theory thus predicts that proteins have a lower bound on the size required to achieve discrimination, and that they will be larger than this in order to be evolvable (\fref{fig:fig6}A). It is not clear how close proteins get to this minimum, but it would be more likely in prokaryotes since they have greater efficiency requirements.
  \\

  \nsb{Examining and predicting experimental trends}
    Affinity and specificity vary non-monotonically with $\dlig$, $K$, and $\epsilon$,
  so there are no general, unidirectional trends. Still, we can discuss trends found in the experimental literature in the context of \fref{fig:fig5}A-B, and offer explanations that are consistent with our model and lead to testable predictions.
  
  Many studies report that germline antibodies are flexible, and become more rigid in a process known as affinity maturation.~\cite{vanjm14,misfi18} A recent \textit{in silico}
  directed evolution study corroborated this, and also showed that some antibodies
  first become more flexible, before later becoming more rigid.~\cite{ovcel18}
  According to our model (cyan arrows, \fref{fig:fig5}A),
  the former case should occur when antibodies have close to optimal shape mismatch
  (low $|\dlp|$); the latter case should occur when antibodies have high shape
  mismatch (high $|\dlp|$). This prediction can be tested by measuring
  shape (and chemistry; note we operationalized mismatch through shape, but shape and chemistry are inseparable in real molecules) mismatch: one can use measures based on static structure,~\cite{cheps03,chepc10,yanbb19} but for better results
  one should calculate how often the optimal binding configuration is sampled in the free antibody,
  using molecular dynamics simulations.~\cite{karns02,munpc12,pabel17,kampn21}
  
  Our model supports the notion that ancestral enzymes were both flexible and
  promiscuous,~\cite{petjt18} as flexibility is often anti-correlated with specificity in our model.
  However, we find that the correlation between flexibility and specificity
  depends on shape and chemical binding energy (\fref{fig:fig5}B).
  This can be illustrated by two case studies that relate conflicting accounts
  of the role of flexibility in enzyme promiscuity: Flexibility and promiscuity are correlated in a
  group of 57 human cytochrome P450 (CYT) enzymes,~\cite{becjc21} while the opposite
  trend is observed in a group of 147 esterases.~\cite{nutjc21}
  In the case of CYT, we know that the binding pocket is quite small ($\dlp<0$),
  so we can infer that the proteins fall to the right of the optimal line (bottom purple arrow, \fref{fig:fig5}B).
  Thus, our model predicts that as rigidity increases, both specificity and affinity
  will eventually decrease as the protein will be too stiff to deform (bottom red arrow, \fref{fig:fig5}B).
  This can be tested by increasing CYT stiffness via directed evolution.~\cite{hilac16}
  In contrast to CYT, the esterases have open active sites ($\dlp>0$), and we know that promiscuity
  is correlated with the volume of the active site,~\cite{barja15,marac18}
  so we can infer that the proteins fall to the left of the optimal line (top purple arrow, \fref{fig:fig5}B).
  In this case, further increasing esterase flexibility via directed evolution should
  reveal that there is an optimal range of flexibility where specificity is maximized (top red arrow, \fref{fig:fig5}B).
  
  Our model explains how affinity and specificity can be either positively~\cite{eatcb95,gaosm21} or negatively~\cite{greti10}
  correlated within a set of proteins, depending on how they differ in shape mismatch
  and flexibility (SI Fig. 7). For example, affinity and specificity are positively correlated in two cases:
  when proteins differ along optimal line (black line, \fref{fig:fig5}A-B), or orthogonal to the
  optimal line. In the former case, deformation energy decreases when
  affinity and specificity increase; in the latter case, deformation energy increases when
  affinity and specificity increase (SI Fig. 7).
  This prediction may be tested by studying the transcription factor Pho4, where
  increased binding affinity of the cognate CACGTG nucleotide sequence was found to
  improve discrimination of the cognate over the non-cognate CACGTT sequence in 210 variants.~\cite{adics21}
  These variants can be studied using molecular dynamics simulations,
  where deformation energy in our model is analogous to the change
  in internal energy of a protein upon binding \cite{reyjm00,kanpc04,pakjc16}.
  
  A challenge in testing these predictions is the vast amount of data that is needed, since
  one needs to measure multiple dimensions for a combination of proteins and ligands. However,
  multiple methods can characterise shape mismatch, flexibility, chemical bond energy, deformation energy,
  and entropy. We advocate combining molecular dynamics simulations
  (which can characterise flexibility and calculate deformation energy) with high-throughput experiments
  (which can measure binding kinetics), and to develop methods to control for orthogonal effects such as
  differences in protein stability or foldability.~\cite{marsc21} Existing public data sets from previous
  experiments present a facile opportunity in this regard.~\cite{namsc12,huapn15,marac18,piace18,becjc21}\\
  
  \nsb{Molecular discrimination by aminoacyl-tRNA synthetases}
  To evaluate the theory that protein size depends on task difficulty, we need
  to know both the relevant non-cognate ligands and necessary binding specificity. In the
  case of aminoacyl-tRNA synthetases (ARSs), we know the relevant ligands
  (the 20-30 proteogenic and non-proteogenic amino acids present in cells),
  and that they have similar \textit{in vivo} concentrations and similar costs
  associated with missense mutations. This presents a natural control,
  such that task difficulty is reduced to a question of discriminability between cognate and
  non-cognate ligands. This is still difficult to evaluate, but we can start by
  using the available experimental data on pairwise binding specificity of ARSs.
  We can rationalise that one out of a pair is easier to recognize if it is smaller
  by a methyl group (steric exclusion), or has an extra hydroxyl group
  (can form more high energy bonds). Thus, it is difficult to discriminate:
  threonine from serine (minus one methyl), isoleucine from valine (minus one methyl),
  phenylalanine from tyrosine (plus one hydroxyl), and alanine from serine (plus one hydroxyl).~\cite{tawfj20}
  
\begin{figure}  \centering
  \includegraphics[width=0.99\linewidth]{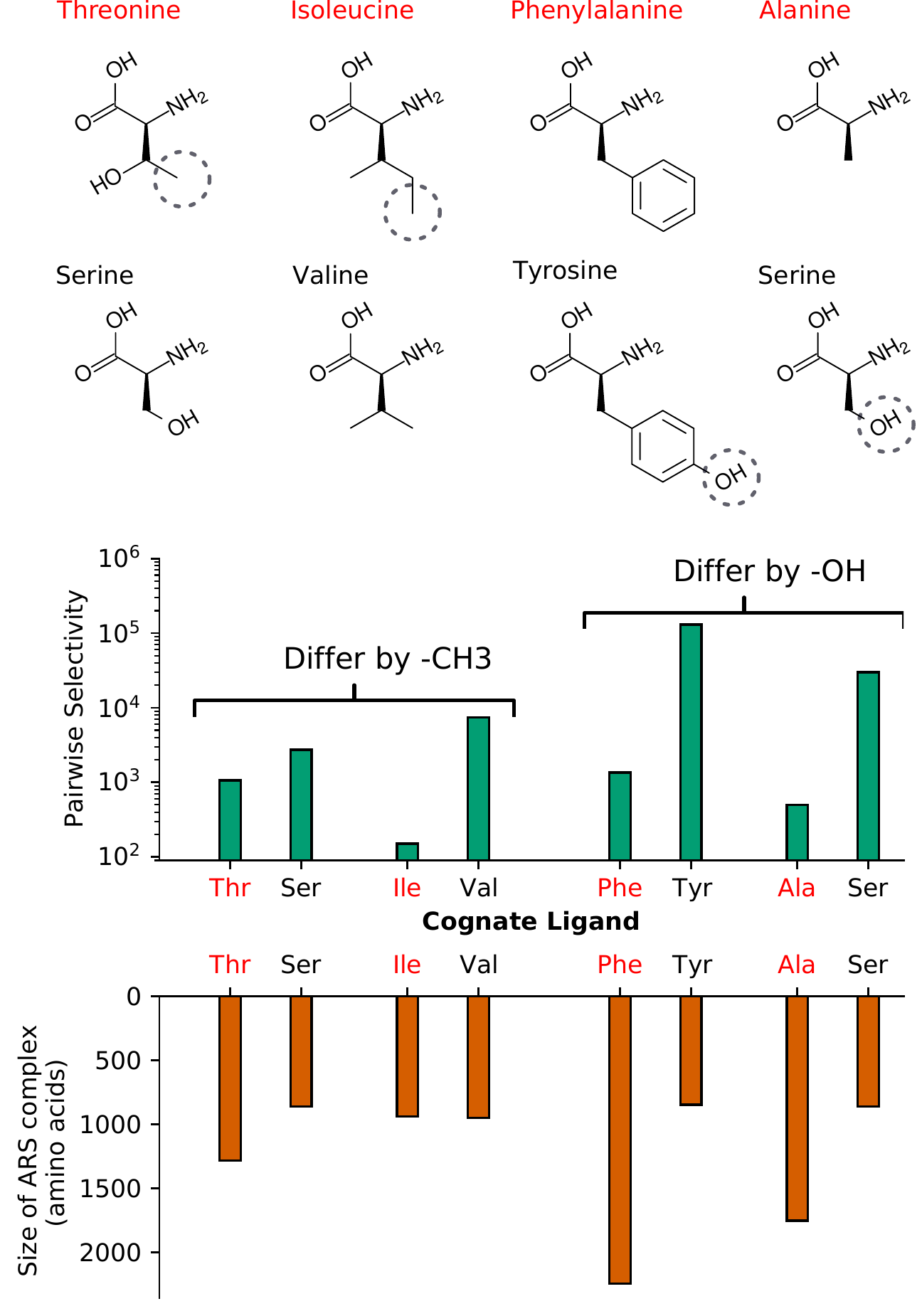}
  \caption{\label{fig:fig7}
  Pairwise selectivity -- ratio of cognate vs non-cognate $K_{\rm M}/ k_{\rm cat}$ -- of aminoacyl-tRNA synthases (ARSs),
  and ARS complex size, of pairs of similar amino acid ligands, where in each pair the labels are coloured according to whether it should be easier (black) or more difficult (red) to discern. Ligands differ by one methyl group (left, ligand with additional $\mathrm{-CH_3}$ in red)
  or one hydroxyl group (right, additional $\mathrm{-OH}$ in black). Differences are highlighted in diagrams (top) with dotted circles.
  }
  \end{figure}

  Comparing ARSs of these pairs, we find that the ARS of the easier-to-recognize ligand has greater specificity and, with the exception of Val-Ile, they are also smaller (\fref{fig:fig7}). This exception may be due to the difficulty in general in distinguishing between many aliphatic amino acids, a point illustrated by the fact that these ARSs all have post-transfer editing domains.~\cite{perbi12} Furthermore, when we compare ARS enzymes to non-ARS enzymes that also act on amino acids (but with lower specificity requirements), the non-ARS
  enzymes tend to be considerably smaller than ARSs (SI Fig. 9).~\cite{tawfj20} These findings support the theory that protein size is related to the difficulty of the discrimination task. We expect that ARSs are an exemplary class with which to further evaluate theories on protein specificity.\\

  \nsb{Limitations and extensions}
  By focusing on binding energy gap we have posed the molecular
  discrimination as a thermodynamic problem. It is not immediately clear whether a kinetics-focused approach would lead to the same conclusions. For example, we find that increasing rigidity can increase specificity, but from a kinetics point of view, one might expect the opposite. There is evidence to suggest that flexibility aids formation of initial encounter complexes, which reduces the rate of futile encounters.~\cite{wanpn13,barja152,planc17} Understanding the role of kinetics in molecular discrimination warrants a separate, focused study.

  We treat chemical binding strength and mismatch (encoded in shape) as separable components in our model. This is useful for studying their relative contributions to molecular discrimination, but molecules cannot be deconstructed in this way. It may be possible, however, to describe the differences in interactions between a protein and several ligands using some type of principal components or reduced dimensions. In this way, one may relate the model results to discrimination of real molecules.
  In a similar vein, we have simplified our model by assuming that ligands are rigid.
  We speculate that if ligands were allowed to move, this would produce results consistent with a rigid ligand and a more flexibile protein. This is consistent with our finding (via the phenomenological model) that many spring constants can be reduced to one effective spring constant. Additionally, if one ligand is more flexible than another, it may be captured by an `effective mismatch' term that takes into account a ligand's ability to change shape.
  In general, protein function is typically governed by a small number of effective parameters, as manifested by drastic dimensional reduction in genotype-to-phenotype map.~\cite{eckbi21} 
  Our simple model demonstrates that specificity is a function of the effective flexibility, shape mismatch, and local chemical interaction,
  but further work on more detailed models is needed to understand how many effective dimensions are needed to describe specificity.

  Our model treats deformations as elastic, so it may not generalize well to proteins that undergo plastic deformations upon binding. In fact, the restructuring of intramolecular bonds can result in a gain in entropy\cite{fenpn12,wanel22}, or an increase in internal enthalpy.~\cite{ferpn07} Furthermore, a simplifying assumption in elastic network models is that bonds are at their equilibrium lengths in the ensemble-average protein structure.  
  However, proteins are non-crystalline matter where many bonds are ``pre-stressed" due to geometrical frustration (the inability to achieve equilibrium length in all bonds simultaneously).~\cite{panfd97,ferqr14}. The frustration gives rise to a rugged energy landscape with myriad of local minima. The model can be easily extended to incorporate frustration by modeling a network with heterogeneous equilibrium lengths $\ell_{ij}$, and considering a thermal ensemble of configurations. Accounting for the role of plasticity and intrinsic disorder in biomolecular discrimination will require a more complex model.\\

  \noindent\sb{Why are proteins so big?~}
  Proteins are large macromolecules. Large proteins require large genomes, and thus slow transcription and translation. Prokaryotic proteins, which prioritize small genomes and fast replication, should benefit immensely from smaller proteins. We do see that
  proteins are on average smaller in prokaryotes (312 residues) than in eukaryotes (441 residues),~\cite{unina18} but they are still quite large. Thus, there is some indispensable functional reason for protein size. We have here proposed that the difficulty of evolving proteins that can discriminate necessitates proteins of a certain size. We now discuss some alternative factors affecting protein size.

  Stability is typically a base requisite for a functional protein,~\cite{sretb84} but most proteins are marginally stable  -- \ie, stability is not maximised, but rather an acceptable level is reached.~\cite{sikjr14} Longer proteins can more easily fulfill these requirements,~\cite{baspe05} but short sequences are often sufficiently stable.~\cite{davns95} Catalytic activity in enzymes is extremely important, but comparable activity levels can be found in much smaller  organic catalysts.~\cite{mausc83,macna08} Large surface area may be needed to make multiple interaction sites, whether for a single molecule,~\cite{henpn20} or multiple molecules.~\cite{paytb83,sretb84,vanjm14} While we propose that size is determined by the difficulty of molecular discrimination, all of the above constraints may also be relevant factors.\\

\nsb{Data availability}
  All code (models, analysis, figures), and some data is available at 
  \href{https://github.com/jomimc/RecProt}{github.com/jomimc/RecProt}.
  Additional data can be shared upon reasonable request.

\nsb{Acknowledgements}
  We thank Jacques Rougemont for discussions. This work was supported by the Institute for Basic Science, Project Code IBS-R020-D1. JPE is partially supported by the Fonds National Suisse Swissmap.

\vskip 0.5cm
\relscale{0.9}

\sb{METHODS}\\

\nsb{Calculating binding entropy}
Binding will typically involve loss of entropy due to constraining the internal fluctuations of the protein. 
To calculate this change in conformational entropy upon binding, one starts by analyzing the elastic energy of a spring network.
  It is given by $\EE = \half \bra{u}  \Hb \ket{u}$~,  where $|u\rangle$ is the displacement vector (of size $d\cdot \NA$) and $\Hb$ is the elasticity matrix (for details see ~\cite{dutpn18,eckrmp19}). Spectral decomposition of $\Hb$ gives $\Hb = \sum_n \lambda_n\dyad{u_n}$, where $\ket{u_n}$ are the normal modes and $\lambda_n$ are the eigenvalues.
  There are $\half d(d+1)=3$ zero energy modes of translation
  and rotation without deformation that we ignore.
  We can then express the energy in normal mode coordinates,
  $\xi_n =\bra{u_n}\ket{u}$, as $\EE = \half \sum_n \lambda_n \xi_n^2$.

  To estimate the entropy, we consider the partition function of the elastic deformations, $\ZZ = \int \dd{\ket{u}} e^{-\beta \EE}$, where $\beta = 1/(kT)$ is the inverse temperature. Expressed in normal coordinates, we get
  $\ZZ = \prod_n {[ \int \dd{\xi_n} \exp(-\beta \lambda_n \xi_n^2/2)]} =\prod_n [2\pi/(\beta \lambda_n)]^{1/2}$,
  where the product is taken over the modes for which $\lambda_n \neq 0$. The entropy  $S$  is obtained from the relation $S = \ln{\ZZ} - \pdv*{\ln{\ZZ}}{\ln{\beta}} $,  so that $S = \half \sum_n {[ \ln (2 \pi e) - \ln (\beta \lambda_n) ]}$. 
  When the protein binds, the modes and their eigenvalues will change. The resulting entropy change is the sum over the changes in the logarithms of the elastic mode energies before and after binding, 
  \be
  \DS = S_{\rm bound} - S_{\rm unbound}= -\half \sum_n {\Delta\ln\lambda_n}~.
  \ee

    Typically, the mode energies $\lambda_n$ increase or stay unchanged upon binding since the motion is more constrained, and therefore the entropy is reduced, $\DS < 0$, as the protein-ligand complex stiffens.~\cite{ricja19} This effect is calculated as follows: 
  We add a new bond at the bottom of the $\Lambdaup$ (opening of the binding site), and make the three bonds between the amino acids of the binding site very rigid by increasing them to $\KL=$ \SI{e4}{kT / nm}$^2$.

  In the model of entropy used in this work, entropy decreases upon binding, and flexible proteins lose more conformational entropy than rigid proteins. This results from our choice of binding stiffness, assuming that binding stiffens the pocket, $\KL = 1000 \geq K$; and that $\KL$ is practically constant for all proteins.
  In reality, flexible proteins will of course have more entropy, and thus more to lose.  Proteins are most often found to lose entropy upon binding, but many proteins instead gain entropy
  due to allosteric conformational change.~\cite{fenpn12} Moreover, there are other contributions to entropy,
  such as solvent entropy and ligand entropy,~\cite{savie08,draeb17,pecao21} that are not treated in our model
  (\eg, contributions of translational and rotational entropy do not depend on the internal degrees
  of freedom and can be excluded).

\nsb{Phenomenological model}
    Consider a free energy $\DG$ which is a function of $r$, the remaining gap in the binding site (see text), and the initial gap $r_0 \approx \ell \dlp$. The elastic energy is a function of the deformation $u  = r_0 - r$, while the chemical energy is a function of the gap $r$, such that the overall free energy is $\DG(r,r_0) = \edf(u) + \ech(r) - T \DS$, with the entropy change $\DS$. For a given mismatch $r_0$ the protein equilibrates at a gap $r=\bar{r}$ when the elastic and chemical forces are equal in magnitude and opposite, $(\pdv*{\DG}{r})_{r_0}= -\edf'(u) + \ech'(r) = 0$. This condition defines an equilibrium line in the $(r,r_0)$-plane,  $M(\bar{r},r_0) = 0$. Taking the differential of the line, $\var{M} = -\edf''(u) \var{r_0} + (\edf''(u) + \ech''(r)) \var{r} =  0$, we obtain its slope, $            \dv*{r_0}{\bar{r}} = 1 + \ech''(r)/\edf''(u)$.
    The binding free energy is the equilibrium value along the line $\bDG(r_0)=\DG(\bar{r},r_0)$. 
    
    The specificity is the free energy difference between a cognate ligand and a competitor, $\DDG  = \bDG(r_0 +\delta r_0) -\bDG(r0) \approx \DD \cdot \delta r_0$, where the ``specific specificity" (or ``discriminabilty") is the specificity per shape difference, $\DD \equiv \dv*{\bDG}{r_0}$. Hence, the specific specificity is exactly the elastic force, $\DD = (\pdv*{\DG}{r_0})_r = \edf'(u)$ (since along the equilibrium, $(\pdv*{\DG}{r})_{r_0}=0$). Maximal specificity is therefore obtained for the mismatch $r_0^\ast$ at which the force is maximal,$\dv*{\edf'(u)}{r_0} =  \edf''(u) (\dv*{u}{r_0}) = \edf''(u)(1 - \dv*{\bar{r}}{r_0}) = 0$.
    It follows that maximal specificity is achieved when the slope of the equilibrium curve is $ \dv*{r_0}{\bar{r}} = 1$. From the equation of the equilibrium line, we find that this optimum is the inflection point of the chemical energy, $\ech''(r) = 0$, where the chemical attractive force is also maximal. For chemical energies that lack an inflection point, one searches for a global maximum of $\ech(r)$. There are also cases with multiple equilibrium points and discontinuities, which need to be considered separately.
     In the text, we apply this general result to the \textbf{MeCh}-model, and obtain the manifold of optimal $r_0^\ast$ mismatch as a function of model parameters $K$, and $\epsilon$.

\nsb{Calculating sequence covariance and correlation}
 We follow the standard procedure: One takes a set of $\NS$ sequences $\mathbf{S}_i$ of length $\NA+\NB$, represented as vectors of ones ($\LS$) and zeros ($\LW$), and subtracts from each the average sequence vector $\mathbf{S}_i \to \mathbf{S}_i - \bar{\mathbf{S}}$ (all entries of $\mathbf{S}_i$ are between 0 and 1). The resulting sequences are the rows of an $\NS \times(\NA+\NB)$ matrix, $\MM$. The covariance matrix is then $\QQ = (\NS-1)^{-1} \MM^\T \MM$. 
 Finally, the matrix $\PP$ of the Pearson correlation coefficients between pairs of positions in the sequence (the ``correlation matrix") is $\PP = \mathrm{dg}\QQ^{-1/2} \QQ \, \mathrm{dg}\QQ^{-1/2}$, where $\mathrm{dg}\QQ$ is the matrix of the diagonal elements of $\QQ$ (\ie, the variances).

\nsb{Connecting genes to structure}
  We allow neighbouring $\LW$ and $\LS$ amino acids to interact such that they either attract, repel, or neither: attraction (repulsion) between two amino acids results in their equilibrium positions moving closer together (further apart) by $\delta \rb = \alpha$ ($\delta \rb = - \alpha$).
  We generate interaction tables $\alpha(a_i, a_j)$ such that $\LW{-}\LW$, $\LW{-}\LS$ / $\LS{-}\LW$, and $\LS{-}\LS$ bonds, respectively result in one of the three possible interactions. Thus, we can generate up to $15$
  unique interaction tables (reduced from a total of $3^4=81$ possible tables, by accounting for symmetries). We show results for one set in \fref{fig:fig4}; we verified that the results do not depend on a particular set, and show results for another set in SI Fig.~10. 
  The new equilibrium position $\rb_i$ of each amino acid is then determined by the original equilibrium position $\Tilde{\rb}_i$ plus the average over the displacements induced by its $N_j$ neighbours,
  \be
    \rb_i = \tilde{\rb}_i + \frac{1}{N_j}\sum_j^{N_j} \alpha(a_i,a_j)\, \hat{\rb}_{ij} ~,
  \ee
  where $j$ is the index of a bonded neighbour and $\hat{\rb}_{ij} =(\tilde{\rb}_j - \tilde{\rb}_i)/|\tilde{\rb}_j- \tilde{\rb}_i|$ is a unit vector pointing from $\tilde{\rb}_i$ to $\tilde{\rb}_j$. The equilibrium lengths of the springs are then set to $\ell_{ij} = |\rb_i - \rb_j|$.
  An alternative approach is to model mutational deformations as the response to linear perturbations of the network's Hamiltonian,~\cite{echcp08,marcr20} as there is mounting evidence of correlations between protein mechanics and structural evolution.~\cite{tanpr21,eckrmp19}

\nsb{The optimization algorithm}
  For fixed coupling constants, we find a configuration which optimizes energy. After moving the protein close to the ligand, we use the L-BFGS method~\cite{byrsj95} to find a minimum. To model steric repulsion, we restrict the  positions of the amino acids so that they are outside the sector defined by the ligand position and angle $\theta$. Two post-processing checks precede the analysis: We require that the orientation of each triangle in the protein maintains its orientation, \ie, that the surface is not flipped, and the minimized energy must be at least as low as that obtained by a completely rigid protein. We discard any results that do not pass these tests.

\nsb{Evolvability and robustness}
  \textit{Evolvability} is the ability of a population of organisms to evolve new phenotypes. We measure this as the maximum number of mutations a fit sequence can accumulate without having fitness less than zero. \textit{Robustness} is the ability of a sequence to mutate while retaining its original function. We measure robustness as the minimum number of mutations needed for a fit sequence to become non-fit. In practice, we calculated the edit distance between fit sequences, and clustered them using a density-based clustering algorithm (DBSCAN, implemented in the sklearn python module; $\mathrm{eps}=1$, min\_samples $=1$).~\cite{pedjm11} Evolvability of a sequence is then the maximum distance within a cluster, while robustness is the minimum distance within a cluster. For both measures we averaged over all fit sequences (according to  \eref{eq:fit}), and normalized by dividing by the genome size.

\end{document}


\title{Supporting information for \\ ``General theory of specific binding: \\ insights 
from a genetic-mechano-chemical protein model''}

 \author{John M. McBride}
   \email{jmmcbride@protonmail.com}
   \affiliation{Center for Soft and Living Matter, Institute for Basic Science, Ulsan 44919, South Korea}
 \author{Jean-Pierre Eckmann} 
   \email{Jean-Pierre.Eckmann@unige.ch}
   \affiliation{D\'{e}partement de Physique Th\'{e}orique and Section de Math\'{e}matiques, University of Geneva, Geneva, Switzerland}
 \author{Tsvi Tlusty}
  \email{tsvitlusty@gmail.com}
   \affiliation{Center for Soft and Living Matter, Institute for Basic Science, Ulsan 44919, South Korea}
   \affiliation{Departments of Physics and Chemistry, Ulsan National Institute of Science and Technology, Ulsan 44919, South Korea}

\maketitle

  \clearpage

  \begin{figure}[t!]  \centering
  \includegraphics[width=0.70\linewidth]{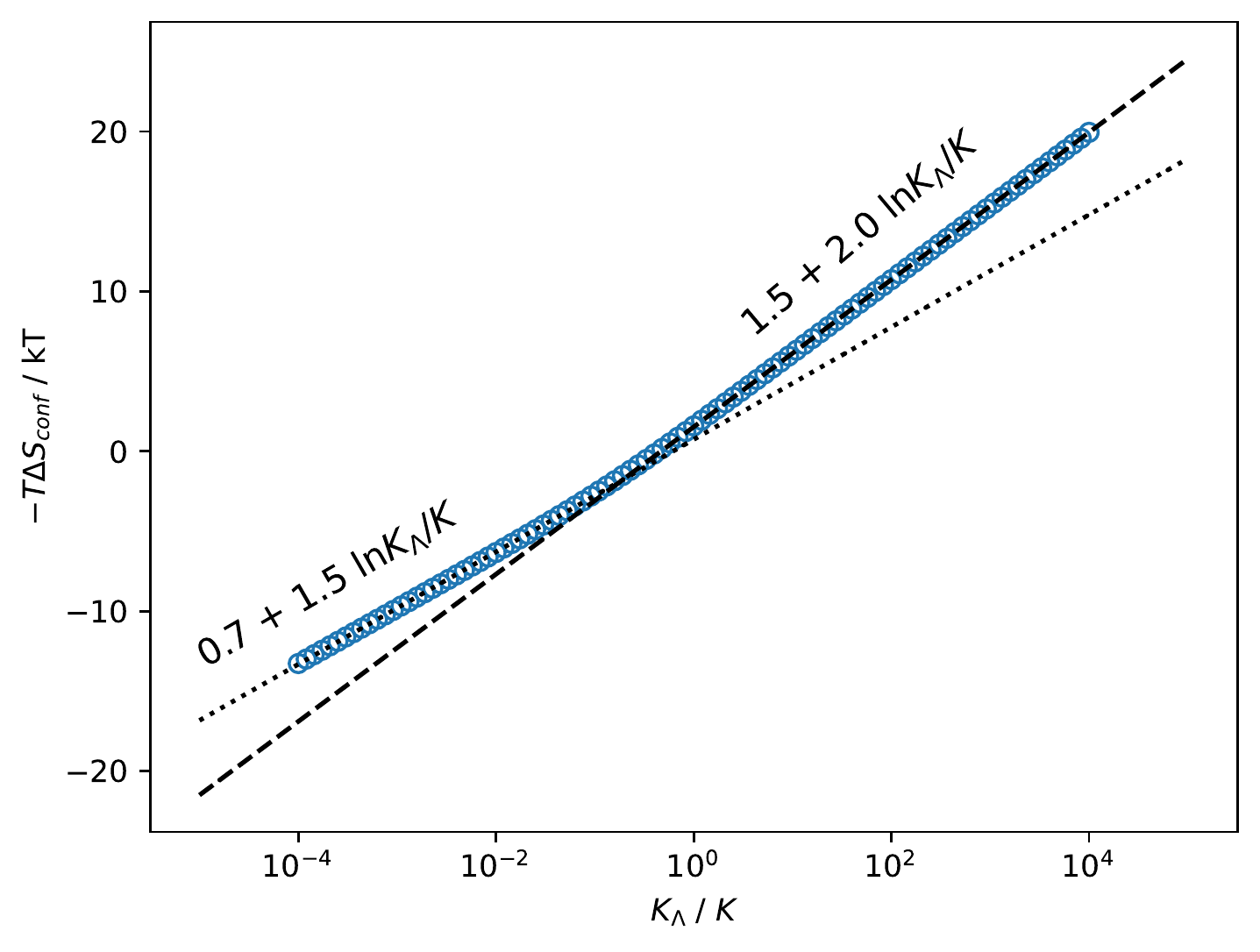}
  \caption{\label{fig:si8}
  Scaling of conformational entropy change upon binding, $-T\DSC$, with spring constant, $K$, for the \textbf{MeCh} model.
  Change in conformational entropy is calculated for protein size "w2h1", and the stiff springs made between the protein and the ligand have spring constant $\KL =\SI{e4}{kT/\nm^{2}}$.
  }
  \end{figure}

  \clearpage

  \begin{figure}[t!]  \centering
  \includegraphics[width=0.99\linewidth]{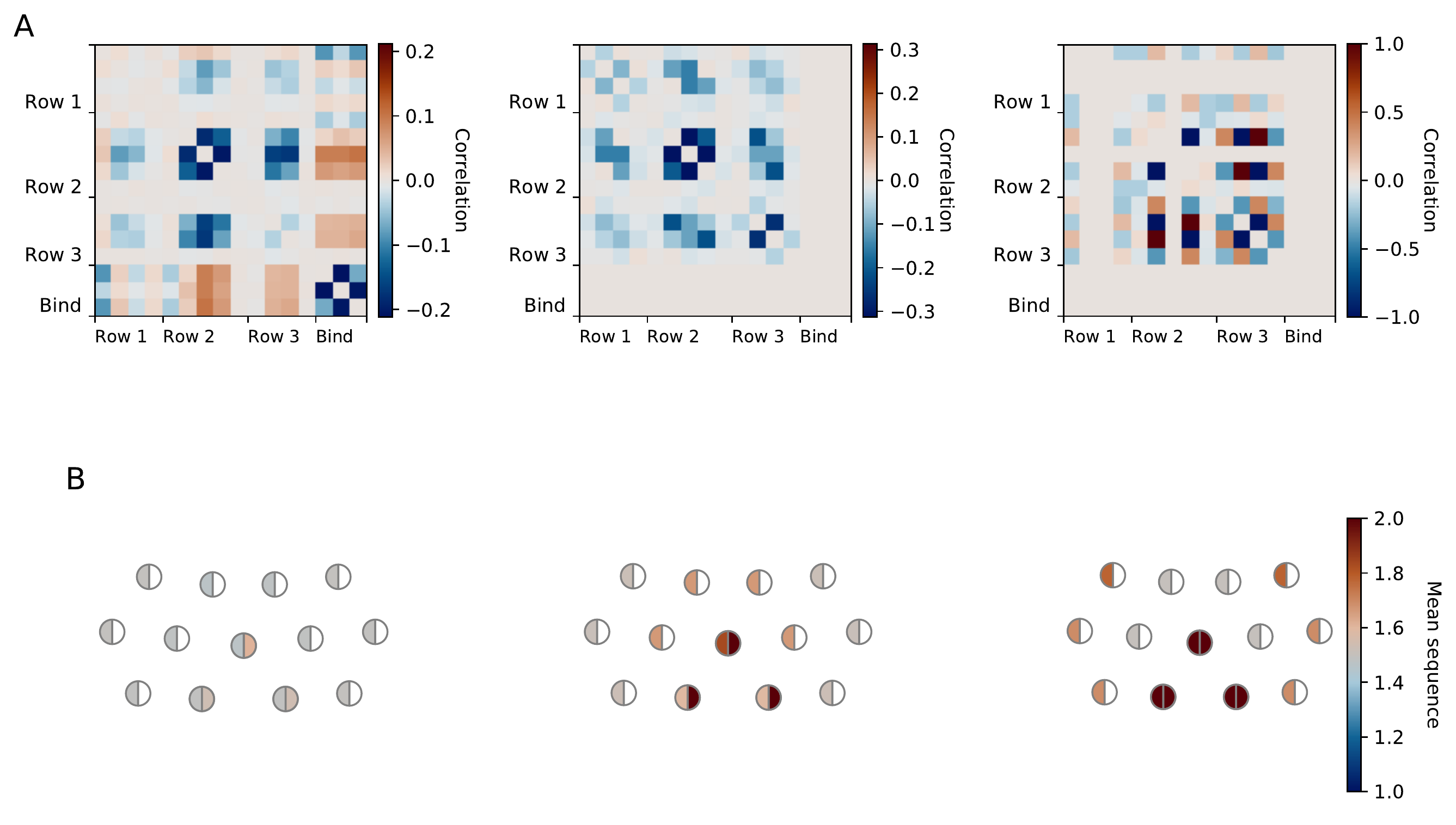}
  \caption{\label{fig:si1}
  A-B: Sequence correlation (A) and average sequence (1 if all weak, 2 if all strong) (B) for all loose-recognition solutions (left),
  for solutions with $\DDG >\SI{3.6}{kT}$ (middle), and for solutions with 
  $\DDG > \SI{3.8}{kT}$ (right). Left-filled semicircles indicate the sequence element (codon) that determines the spring constant; the three right-filled circles at the binding site indicate
  the sequence element that determines binding strength.
  For this example (same as in the main text), the parameters are:
  $K_s=1000$, $K_m=400$, $K_w=\SI{100}{kT/\nm^{2}}$;
  $\epsilon_s = 6$, $\epsilon_w = 1.75$ kT; $\thetS=\ang{80}$, $\thetL=\ang{90}$. \\ \\
  Out of all the loose-recognition solutions, there is a slight tendency for weaker springs,
  and stronger chemical binding, compared to a random set of sequences.
  For solutions with $\DDG > \SI{3.6}{kT}$, all of the sequences have maximal chemical interaction
  strength ($\LLS$), and they tend to have a rigid core.
  For solutions with $\DDG > \SI{3.8}{kT}$, all of the sequences have maximal chemical interaction strength ($\LLS$), and maximally rigid binding site; in this case, the exterior tends to be rigid rather than the core.
  }
  \end{figure}

  \clearpage

  \begin{figure}[t!]  \centering
  \includegraphics[width=0.4\linewidth]{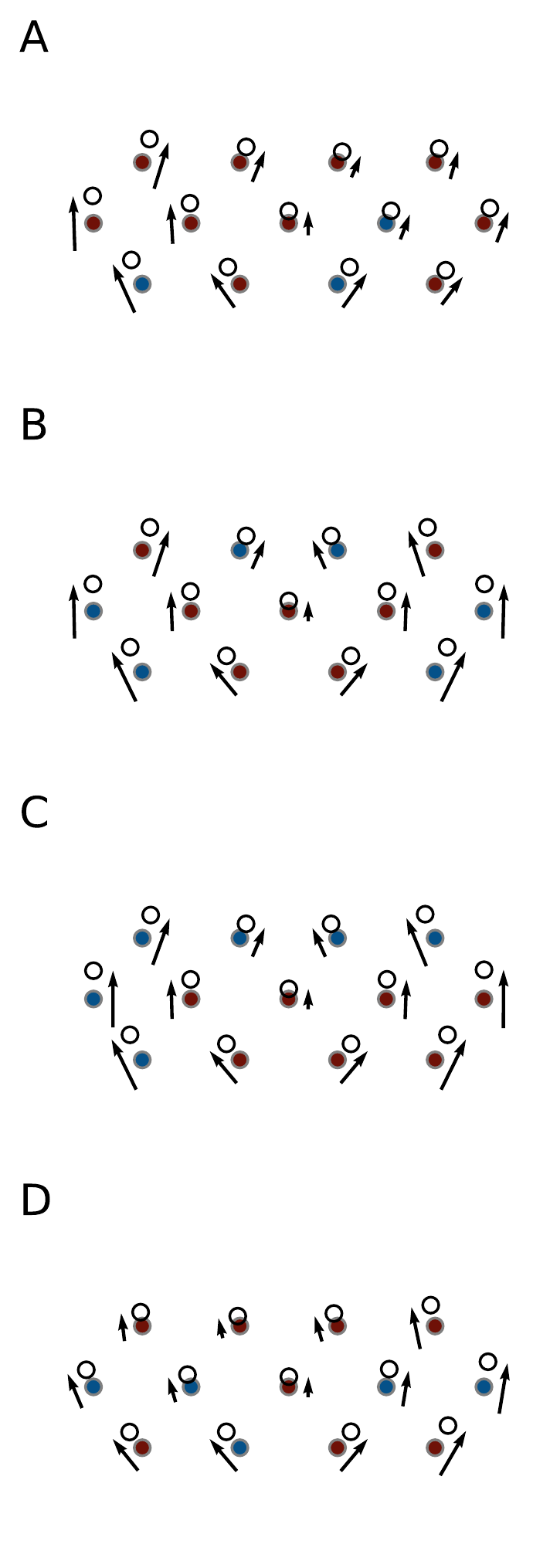}
  \caption{\label{fig:si1b}
  A-D: Four examples of solutions with $\DDG > 3.6$ kT. The coloured circles are in the equilibrium positions, and they are coloured according to sequence. The arrows indicate the relative magnitude, and direction, of displacement upon binding, and the open circles show the bound configuration.
  \\ \\
  Note that many different arrangement of weak / strong springs lead to a solution to the discrimination problem. Ultimately, it is only the resulting effective flexibility at the binding site that matters, and the exact fraction of strong / weak springs, or their specific distribution, is not of special importance.
  }
  \end{figure}

  \clearpage

  \begin{figure}[t!]  \centering
  \includegraphics[width=0.90\linewidth]{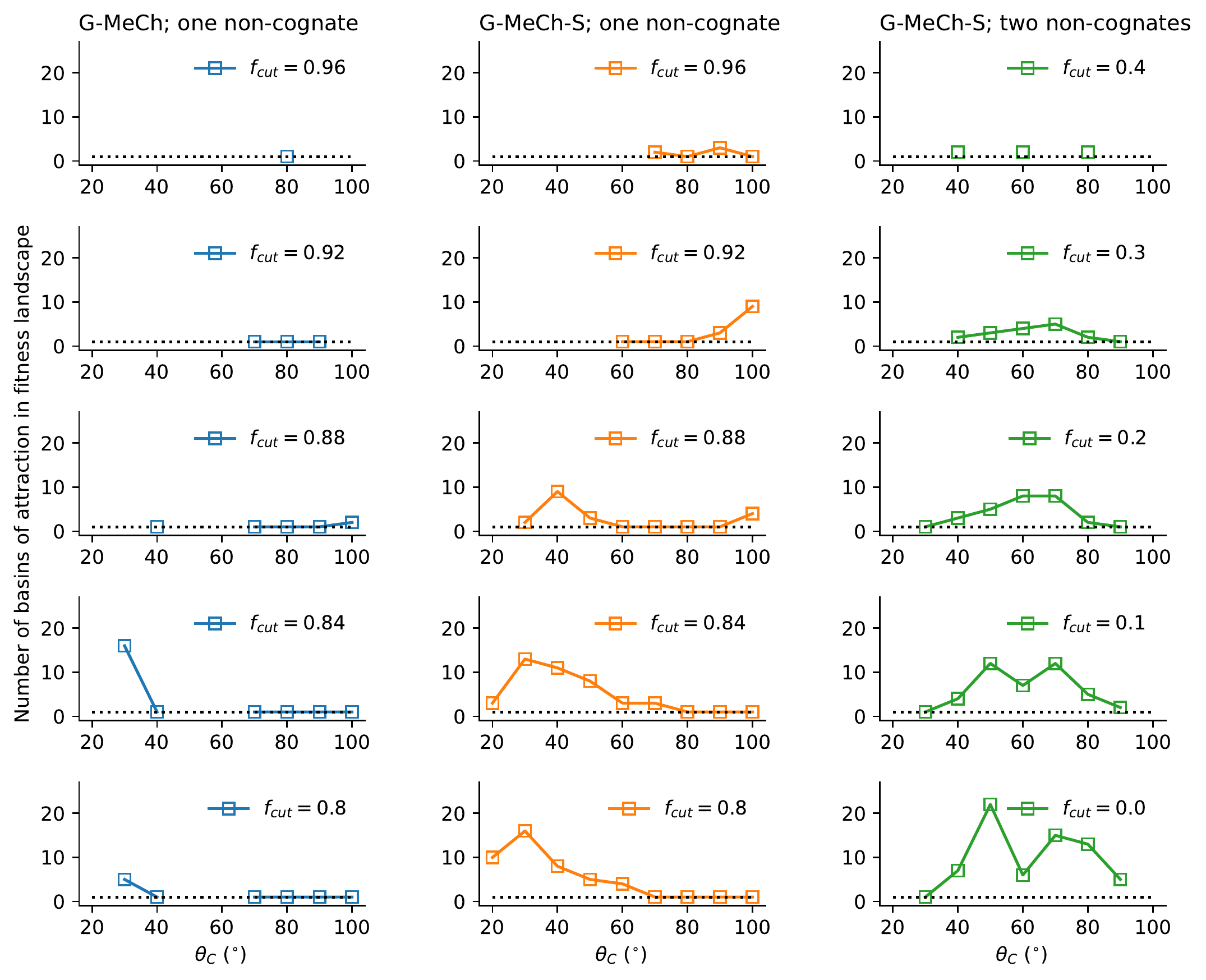}
  \caption{\label{fig:si9}
  Number of basins of attraction in fitness landscapes as a function of $\theta_C$, the cognate ligand, for three types of fitness landscape:
  \textbf{G-MeCh} model (left), with fitness defined as using only a single non-cognate ligand; \textbf{G-MeCh-S} model (middle), with fitness defined the same way;
  \textbf{G-MeCh-S} model (right), with fitness defined using two non-cognate ligands (as is done in the main text). Each subplot shows the number of basins (cluster of `fit' sequences that can all be reached via point mutations) for different values of fitness cutoff $f_{cut}$. The black dotted line indicates a single basin; as $F_\mathrm{cut}$ decreases, each set of data should approach this line as all sequences will be fit. \\
  The equilbrium structure of the unbound protein is fixed in the \textbf{G-MeCh} model. As a result, the mismatch and therefore the fitness landscape are quite smooth functions of the sequence, so we typically only see a single basin of attraction in the landscape.
  In \textbf{G-MeCh-S} model, by  adding structural perturbations, we create roughness in the fitness landscape, resulting in rugged landscape with multiple basins.
  We find that increasing task difficulty by increasing the number of non-cognate ligands (rather than, \eg, changing the shape difference between the cognate and  non-cognate ligands) results in a more tractable fitness landscape. The fraction of fit sequences is not so sensitive to the value of $F_\mathrm{cut}$, so the subsequent calculations of evolvabililty and robustness are computationally feasible (these calculations become memory intensive for large numbers of sequences).
  }

  \end{figure}



  \clearpage

  \begin{figure}[t!]  \centering
  \includegraphics[width=0.70\linewidth]{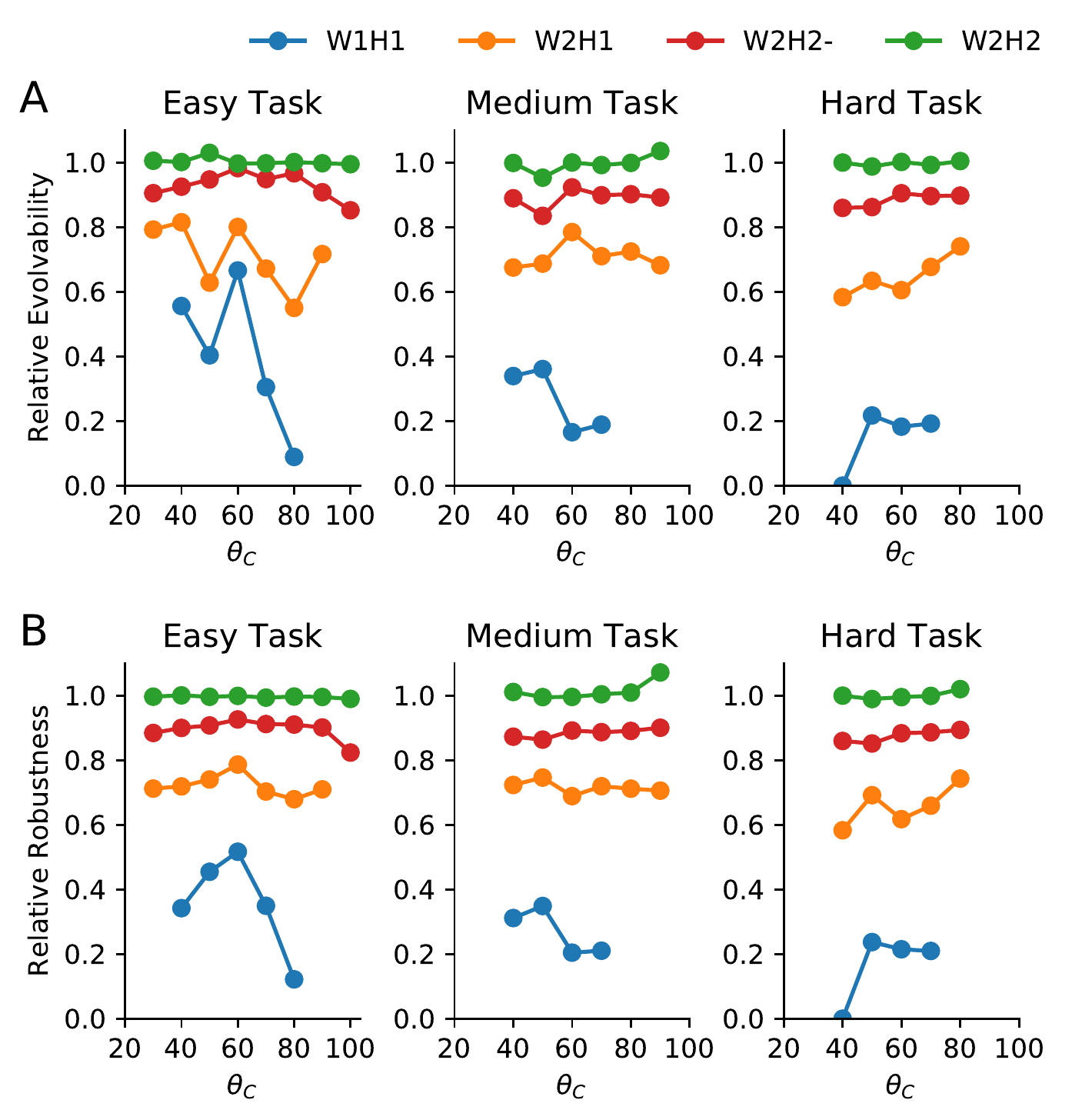}
  \caption{\label{fig:si2}
  Evolvability (A) and robustness (B) relative to the largest protein (W2H3-),
  as a function of the cognate ligand shape, $\theta_C$ (degrees)
  for easy ($\phi=\ang{10}$), medium ($\phi=\ang{20}$) and hard ($\phi=\ang{30}$) tasks.
  The task difficulty, $\phi$, determines the number of non-cognate ligands that contribute negatively to fitness: non-cognate ligands negatively affect fitness if $|\theta_C - \theta_{NC}| \leq \phi$.
  Colours indicate protein models of different size.
  }
  \end{figure}

  \clearpage

  \begin{figure}[t!]  \centering
  \includegraphics[width=0.95\linewidth]{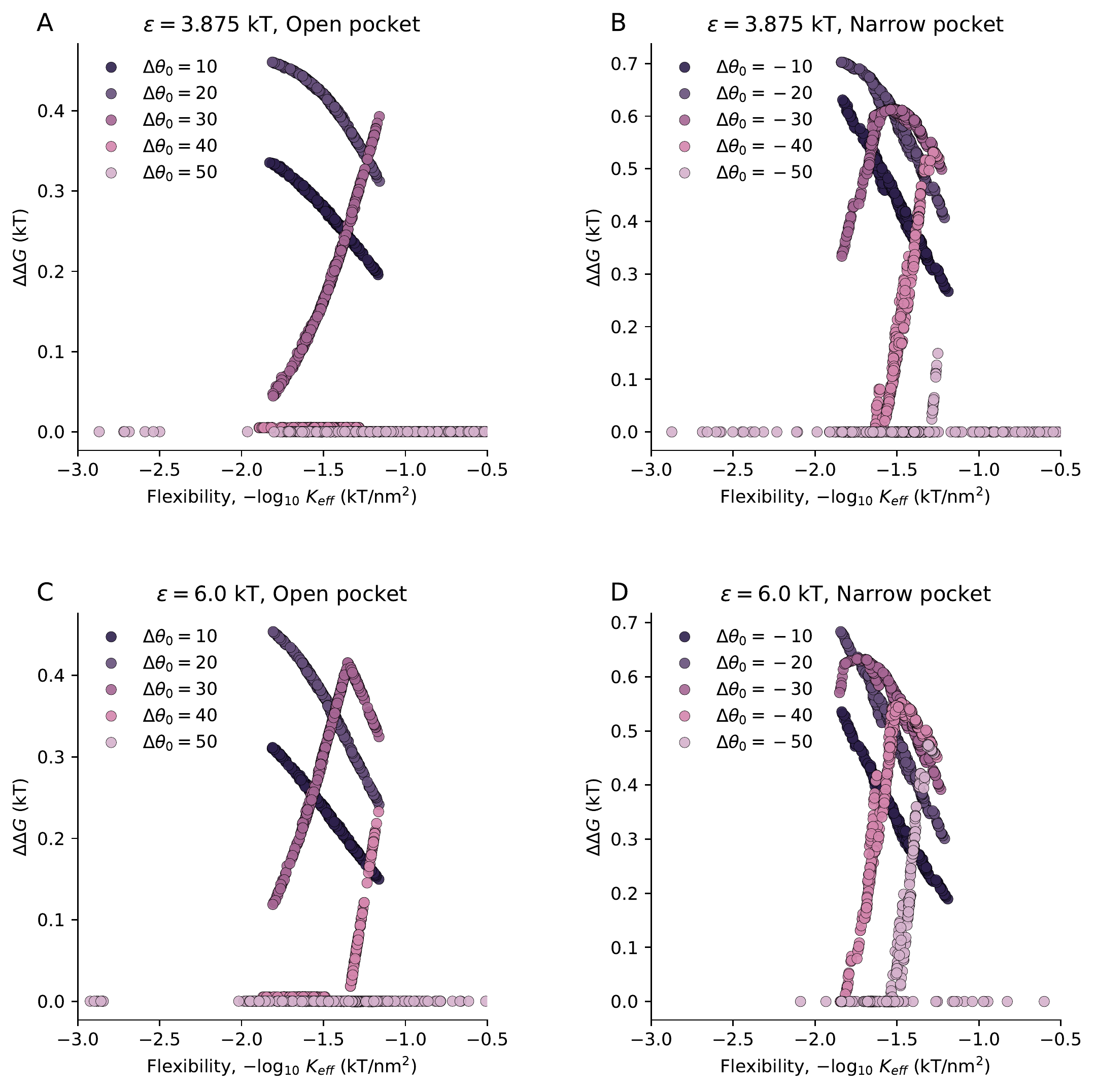}
  \caption{\label{fig:si10}
  Specificity $\DDG$ as a function of the effective spring constant $\Keff$ for a heterogeneous
  spring network (\textbf{G-MeCh} model), for open (A, C) and narrow (B, D) binding pockets,
  for different values of chemical binding energy $\epsilon$ (A-B, weak; C-D, strong),
  for different values of shape mismatch $\dlig$.
  For this example, the parameters are: $\Ks=\num{e3}$, $\Km=\num{400}$, $\Kw=\num{100}$, $\KL=\SI{e4}{kT/\nm^2}$; $\epss = 6$, $\epsw = \SI{3.88}{kT}$; $\thetS=\ang{80}$, $\thetL=\ang{90}$; $T\DSO=\SI{0}{kT}$. \\
  $\Keff$ is obtained by measuring the displacement $r$, $r=\sin (\theta_0 - \theta_{\mathrm{equil}})$ (where $\theta_{\mathrm{equil}})$ is the angle of the protein binding pocket in its bound equilibrium configuration), of one of the three protein binding sites
  (assuming the other two are not displaced), and assuming that this displacement is stretching a single spring: $\Keff = 2\EE/r^2$. \\
  The results show exactly the same trends as the specificity phase diagram for the \textbf{MeCh} model:
  At low mismatch, specificity is aided by low flexibility. As mismatch is increased, there is an optimum, intermediate degree of flexibility. As mismatch is increased further, the specificity tends to zero for all degrees of flexibility.
  }
  \end{figure}

  \clearpage

  \begin{figure}[t!]  \centering
  \includegraphics[width=0.95\linewidth]{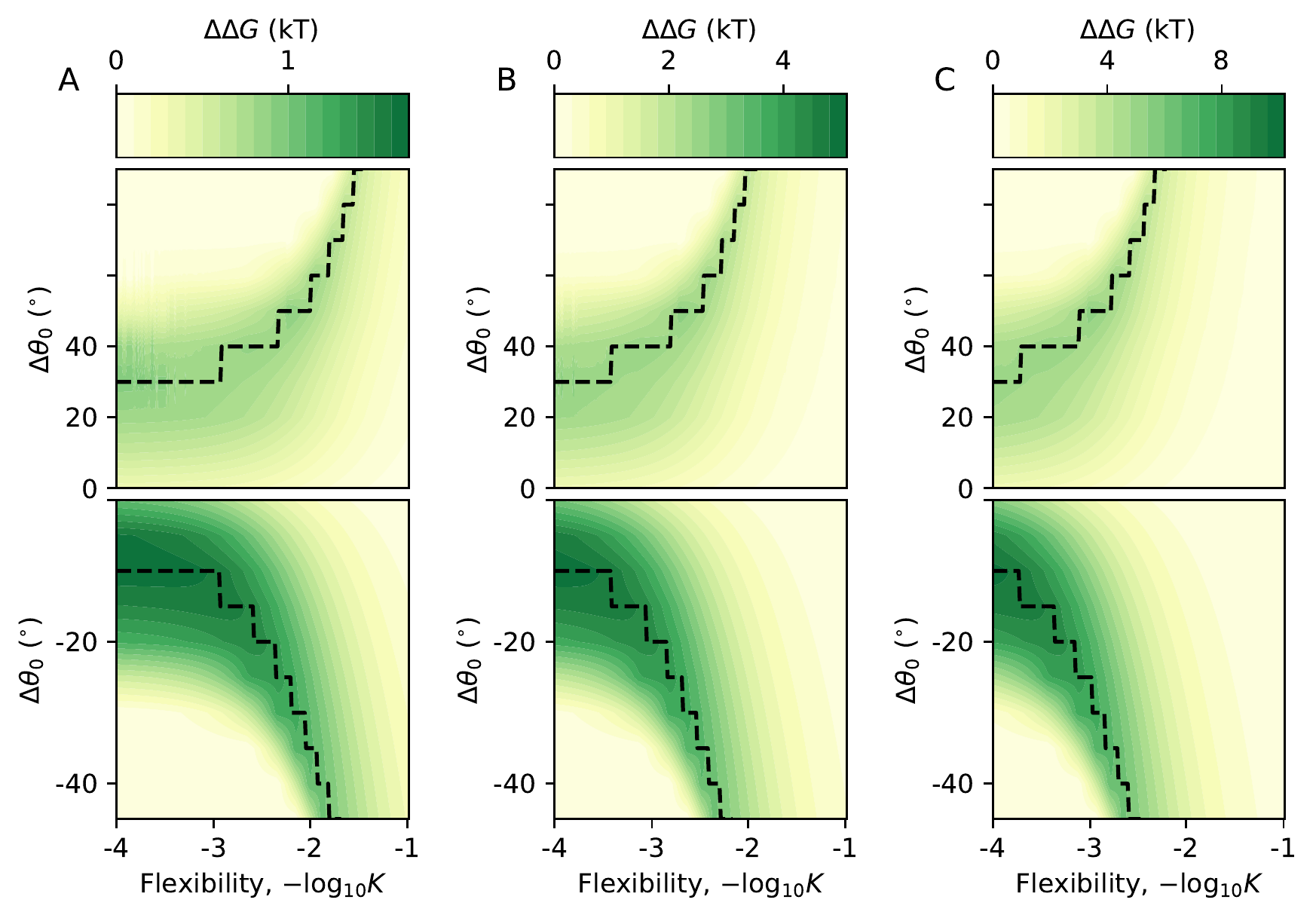}
  \caption{\label{fig:si11}
  Specificity $\DDG$ as a function of protein flexibility,
  and shape mismatch between cognate ligand and binding pocket
  ($\epsilon=\SI{8}{kT}$, $|\dlig|=\ang{10}$, $-T\DSO=0$, $\KL=\SI{e4}{kT/\nm^2}$).
  Separate plots are shown for $\epsilon = \SI{2}{kT}$ (A), $\epsilon = \SI{6}{kT}$ (B), and
  $\epsilon = \SI{12}{kT}$ (C).
  In the top plots the cognate ligand is smaller than
  the non-cognate, and the binding pocket is open to the ligand; in the bottom plots the cognate
  ligand is larger than the non-cognate, and the binding pocket is narrow compared to the ligand.
  The optimal mismatch for a given flexibility is shown by the black line.
  \\ \\
  Here we can see that the shape of the specificity diagram does not change with $\epsilon$, but the optimal line is shifted,
  in line with the $\epsilon/K$ scaling predicted by the phenomenological model.
  }
  \end{figure}

  \clearpage

  \begin{figure}[t!]  \centering
  \includegraphics[width=0.95\linewidth]{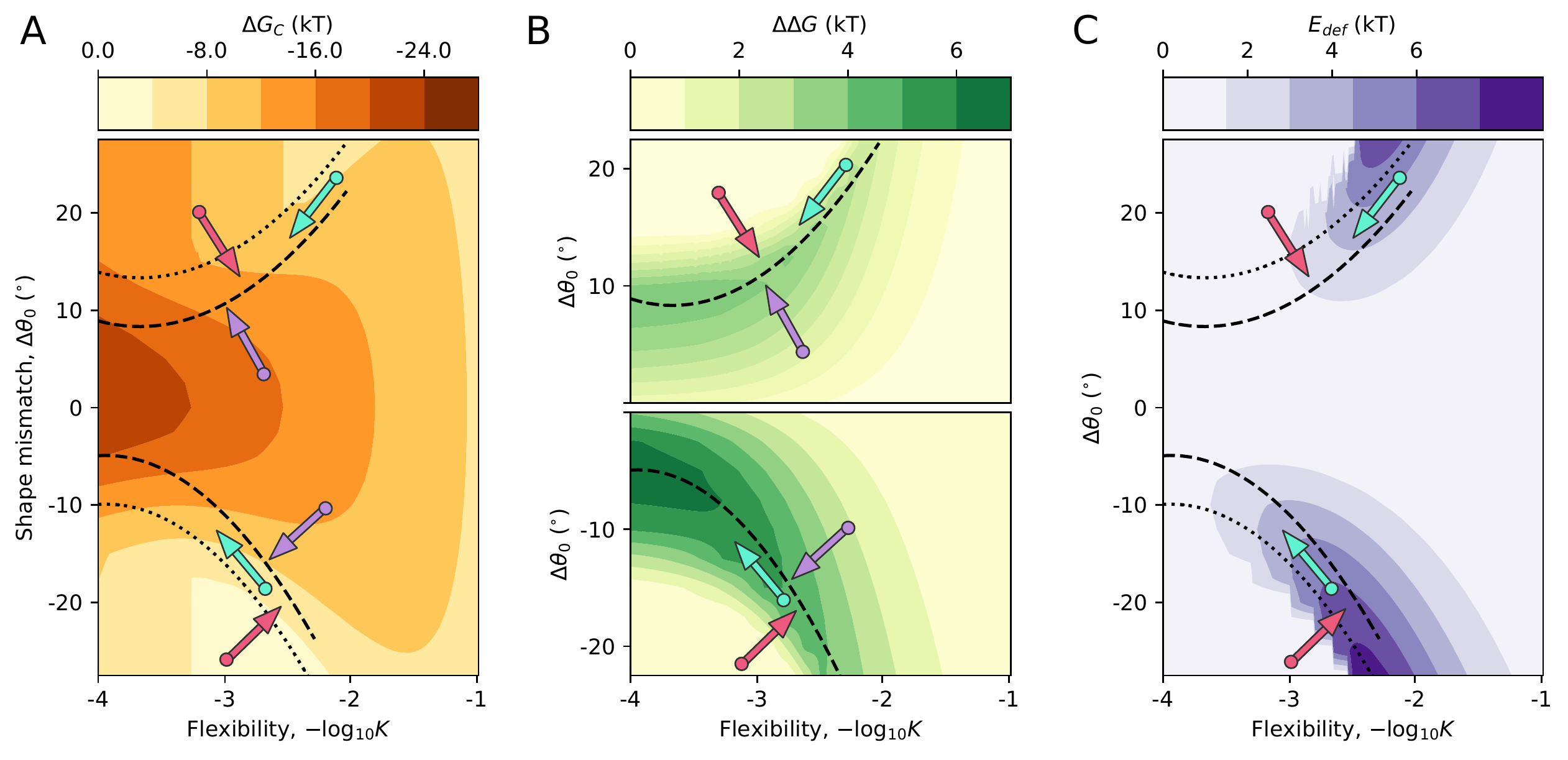}
  \caption{\label{fig:spec_aff_def}
  Binding affinity (A), specificity (B) and deformation energy (C) as a function of protein
  flexibility, and shape mismatch between cognate ligand and binding pocket
  for the \textbf{MeCh} model
  ($\epsilon=\SI{8}{kT}$, $|\dlig|=\ang{10}$, $-T\DSO=0$, $\KL=\SI{e4}{kT/\nm^2}$).
  In the top plot (B) the cognate ligand is smaller than
  the non-cognate, and the binding pocket is open to the ligand; in the bottom plot (B) the cognate
  ligand is larger than the non-cognate, and the binding pocket is narrow compared to the ligand.
  The optimal mismatch for a given flexibility (\ie, with maximal specificity) is shown by the dashed line;
  the optimum for the non-cognate ligand is shown by the dotted line.
  Red arrows indicate regions where affinity is positively correlated with specificity
  and deformation energy increases with specificity.
  Cyan arrows indicate regions where affinity is positively correlated with specificity
  and deformation energy decreases with specificity.
  Purple arrows indicate regions where affinity is negatively correlated with specificity.
  }
  \end{figure}

  \clearpage

  \begin{figure}[t!]  \centering
  \includegraphics[width=0.70\linewidth]{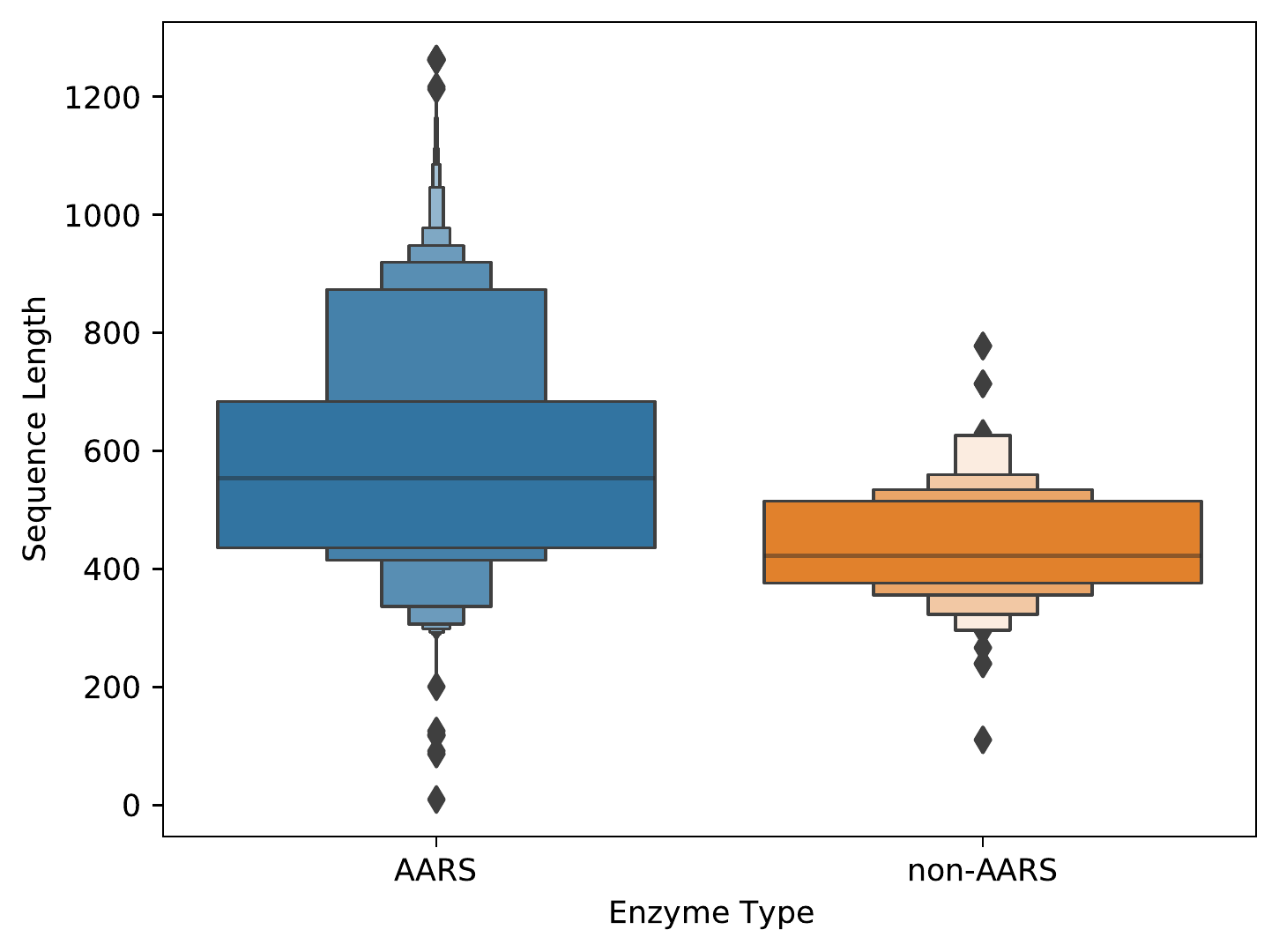}
  \caption{\label{fig:si4}
  Sequence length distributions of enzymes whose substrates are
  amino acids. We show separate distributions depending on
  whether the enzyme is an aminoacyl-tRNA synthetase (AARS) or not.
  }
  \end{figure}

  \clearpage

  \begin{figure}[t!]  \centering
  \includegraphics[width=0.70\linewidth]{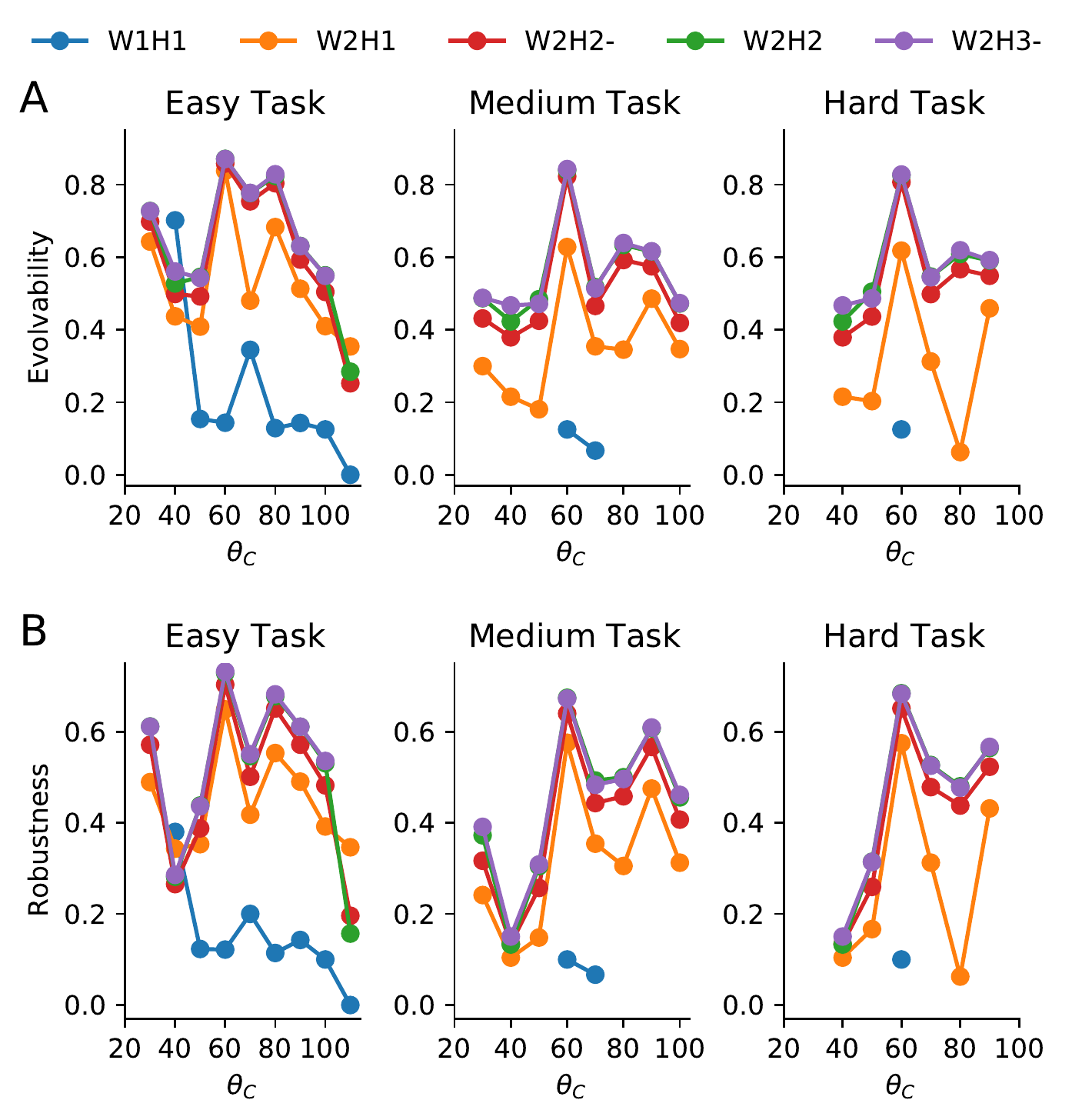}
  \caption{\label{fig:si5}
  Evolvability (A) and robustness (B) as a function of the angle of
  the cognate ligand, $\theta_C$, for easy, medium, and hard recognition tasks.
  Colours indicate protein models of different size. Purple and green lines often overlap.
  Symbols are only shown for ligands for which solutions were found.
  These results were obtained for structural perturbations resulting from $\delta r=0.1$,
  according to sequence: $\LW \LW$, $-\delta r$; $\LW \LS$, or $\LS\LW$, $\delta r$;
  $\LS \LS$, $\delta r$.
  }
  \end{figure}